\documentclass{egpubl}
\usepackage{eg2025}

\usepackage[T1]{fontenc}
\usepackage{dfadobe} 

\biberVersion
\BibtexOrBiblatex
\usepackage[backend=biber,bibstyle=EG,citestyle=alphabetic,backref=true]{biblatex} 
\electronicVersion
\PrintedOrElectronic

\ifpdf \usepackage[pdftex]{graphicx} \pdfcompresslevel=9
\else \usepackage[dvips]{graphicx} \fi

\usepackage{egweblnk} 

\usepackage[ruled]{algorithm2e} 

\usepackage{multirow}
\usepackage{amsmath}
\usepackage{amssymb}
\usepackage{booktabs}

\usepackage{color}
\definecolor{olive}{rgb}{0.1,0.8,0.3}
\definecolor{mauve}{rgb}{0.48,0,0.72}

\usepackage[defaultcolor=red,commandnameprefix=ifneeded]{changes}

\title[DragPoser: Motion Reconstruction from Variable Sparse Tracking Signals via Latent Space Optimization]{DragPoser: Motion Reconstruction from Variable Sparse Tracking Signals via Latent Space Optimization}

{
\author[J. L. Ponton et al.]
{\parbox{\textwidth}
    {\centering 
    J. L. Ponton$^{1}$\orcid{0000-0001-6576-4528}
    E. Pujol$^{1}$\orcid{0000-0001-6406-7499}
    A. Aristidou$^{2,3}$\orcid{0000-0001-7754-0791}
    C. Andujar$^{1}$\orcid{0000-0002-8480-4713}
    and N. Pelechano$^{1}$\orcid{0000-0002-1437-245X}
    }
    \\
    {\parbox{\textwidth}
        {\centering
        $^1$Universitat Politècnica de Catalunya, Barcelona, Spain\\
        $^2$University of Cyprus, Nicosia, Cyprus\\
        $^3$CYENS Centre of Excellence, Nicosia, Cyprus
        }
    }
}

\begin{document}

 \teaser{
  \includegraphics[width=1.0\linewidth]{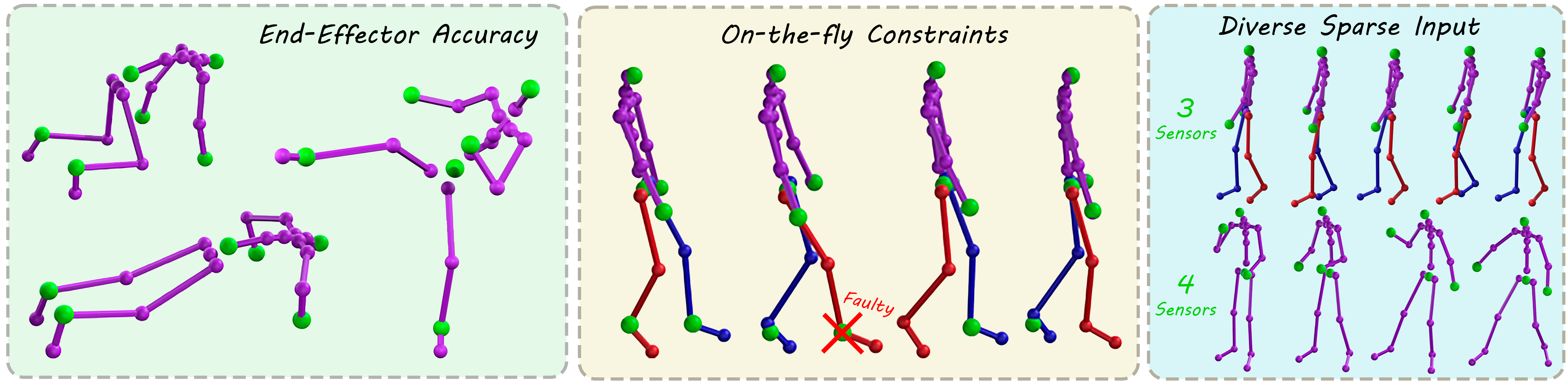}
  \centering
   \caption{Our approach reconstructs complete full-body poses using only a few sensors (highlighted in green). Our technique achieves high-quality animation while precisely locating end-effector constraints (left). Furthermore, it seamlessly incorporates additional real-time constraints, enabling handling scenarios such as missing sensor data (middle). Notably, our versatile one-for-all architecture can adapt to arbitrary combinations of sensors (right), eliminating the need to modify the network's architecture or retrain.}
 \label{fig:teaser}
}

\maketitle

\begin{abstract}
    High-quality motion reconstruction that follows the user's movements can be achieved by high-end mocap systems with many sensors. However, obtaining such animation quality with fewer input devices is gaining popularity as it brings mocap closer to the general public. The main challenges include the loss of end-effector accuracy in learning-based approaches, or the lack of naturalness and smoothness in IK-based solutions. In addition, such systems are often finely tuned to a specific number of trackers and are highly sensitive to missing data, e.g., in scenarios where a sensor is occluded or malfunctions. In response to these challenges, we introduce DragPoser, a novel deep-learning-based motion reconstruction system that accurately represents hard and dynamic constraints, attaining real-time high end-effectors position accuracy. This is achieved through a pose optimization process within a structured latent space. Our system requires only one-time training on a large human motion dataset, and then constraints can be dynamically defined as losses, while the pose is iteratively refined by computing the gradients of these losses within the latent space. To further enhance our approach, we incorporate a Temporal Predictor network, which employs a Transformer architecture to directly encode temporality within the latent space. This network ensures the pose optimization is confined to the manifold of valid poses and also leverages past pose data to predict temporally coherent poses. Results demonstrate that DragPoser surpasses both IK-based and the latest data-driven methods in achieving precise end-effector positioning, while it produces natural poses and temporally coherent motion. In addition, our system showcases robustness against on-the-fly constraint modifications, and exhibits adaptability to various input configurations and changes. The complete source code, trained model, animation databases, and supplementary material used in this paper can be found at \href{https://upc-virvig.github.io/DragPoser}{https://upc-virvig.github.io/DragPoser}


\end{abstract}

\newpage
\section{Introduction}
\label{sec:introduction}


Human motion reconstruction and editing have gained significant attention in diverse industries such as entertainment, sports, and rehabilitation, with applications spanning virtual reality (VR), augmented reality (AR), training, education, simulations, and collaborative work. In professional settings, motion reconstruction typically relies on high-quality optical motion capture (mocap) systems or Inertial Measurement Unit (IMU)-based systems equipped with numerous sensors. Although effective, these systems come with significant costs and require extensive calibrations and technical expertise. Consequently, recent trends indicate a growing demand for cost-effective and minimally intrusive mocap systems that leverage consumer-grade hardware for motion reconstruction. This shift in focus aims to cater to applications in the Metaverse, targeting affordability and accessibility. Motion editing often involves the use of non-learned inverse kinematics (IK) methods such as CCD \cite{kenwright:2012} or FABRIK \cite{aristidou:2011}. While IK enables the rapid creation of animations using end-effectors, it frequently leads to time-incoherent or unnatural human motion due to the inherently underdetermined nature of the problem.


With the growing availability of extensive repositories of motion capture data and the rapid advancements in deep learning, various data-driven methodologies have emerged to grapple with the intricacies and challenges of human motion reconstruction. Recent studies have focused on the reconstruction of poses from sparse input~\cite{ponton2023sparseposer, yi:2023, Winkler:2022}, aiming to reduce reliance on expensive motion capture suits. These methodologies, leveraging their learning from vast and precise mocap datasets, reconstruct human motion that is characterized by both temporal coherence and naturalness, overcoming the shortcomings of traditional IK~\cite{oreshkin2021a, Agrawal:2023}. However, they often encounter limitations arising from their rigid neural network architectures, restricting users to a specific set of sparse inputs or constraints. Additionally, they are sensitive to faulty sensors or tracking losses. Despite their capacity to reconstruct smooth animations, deep learning-based methods commonly face challenges in dealing with hard constraints, often requiring the fallback to traditional IK methods for refining final poses~\cite{Jiang:2022b, Ponton:2022b}, which can inadvertently compromise the naturalness of the data-driven results.


This paper introduces DragPoser, an innovative deep-learning-based system designed for motion reconstruction. Unlike conventional methods that rely on direct full-body pose prediction from sparse inputs or optimize short motion sequences within a latent space, DragPoser leverages a structured latent space and employs a pose optimization process to represent both hard and dynamic constraints accurately. The key difference lies in utilizing an optimization process within the learned latent space, enabling dynamic guidance in the pose search. The system undergoes a singular training phase on an extensive human motion dataset. Subsequently, constraints are dynamically defined as losses, and the pose is iteratively refined by computing gradients within the latent space. To ensure temporal coherence and constrain pose optimization within the valid pose manifold, we integrate a Temporal Predictor network, which directly encodes temporality into the latent space. This departure from traditional direct pose prediction methods enables a more precise and versatile exploration of poses, providing increased generality and improved accuracy when enforcing rigid constraints on end effector positions.

We extensively tested and compared our method against state-of-the-art methods in a variety of challenging motion sequences from public datasets, with different sensor configurations. Our results show that DragPoser outperforms both traditional IK-based methods and the newest data-driven approaches in nearly all error metrics. DragPoser excels at providing accurate end-effector positioning (as in traditional IK systems) but at the same time, it generates natural poses and temporally coherent motion, a feat that data-driven methods typically achieve only in the absence of hard constraints. DragPoser, however, boasts another distinct advantage: its adaptability to hardware changes. We showcase this adaptability by demonstrating DragPoser's robust capability to effectively handle missing input data caused by occlusions, faulty sensors, or communication problems. Additionally, users can dynamically add constraints in real-time to target specific types of motions. This dual adaptability not only reinforces the system's resilience but also enhances its performance in real-world applications. 
\section{Related Work}
\label{sec:related_work}

This section briefly reviews methods utilizing sparse sensor signals to reconstruct full-body poses. We first discuss methods focusing on motion capture from diverse sensor information, and then we focus on data-driven methods overcoming the limitations of traditional inverse kinematics.

\subsection{Motion Capture with Sparse Input}

\paragraph*{Motion Capture with IMU sensors}
Recent research in human motion reconstruction has focused on utilizing fewer Inertial Measurement Units (IMUs) attached to the body. This approach eliminates the need for external sensors or cameras, requires no line-of-sight, and operates seamlessly in diverse environments and lighting conditions. Early methods~\cite{Marcard:2017} used six IMUs for an optimization-based offline pose reconstruction, while later advancements~\cite{Huang:2018, Yi:2021} employed deep learning-based models for real-time accuracy. A key challenge with IMUs is their lack of positional data, leading to inaccuracies in global position estimation. Recent studies~\cite{Jiang:2022}, have introduced Transformer-based models to address this issue, whereas Yi et al.~\cite{Yi:2022} employs physics-based approaches to achieve more realistic motion. However, IMU methods frequently encounter root positional drift over time~\cite{Ami-Williams:2023}. In efforts to mitigate drift, researchers~\cite{guzov:2021, yi:2023, lee:2024} have integrated monocular cameras with SLAM algorithms, aiming for more precise localization.

\paragraph*{Motion Capture with VR 6-DoF Sensors}
The increasing accessibility of commercial VR devices has led to significant progress in reconstructing full-body poses using 6-DoF trackers, which capture both position and rotation. These trackers, utilizing external sensors or computer vision, ensure accurate global information. Early work~\cite{Dittadi:2021} used a variational autoencoder for pose reconstruction from three tracking points, though without global translation estimates. Subsequent studies~\cite{Ahuja:2021, Ponton:2022b} explored matching user poses to a motion dataset, similar to the approach in Motion Matching~\cite{Clavet:2016}. Adding another tracker on the pelvis, Yang et al.~\cite{Yang:2021} developed a model using Gated Recurrent Units (GRUs) to predict lower-body movements with velocity data. To capture the continuous nature of motion, Jiang et al.~\cite{Jiang:2022b} and Zheng et al.~\cite{zheng2023} employed Transformer encoders for real-time pose estimation. Similarly, various generative AI-based methods have been proposed, including conditional flow-based models~\cite{Aliakbarian:2022} and diffusion models~\cite{du2023a, castillo2023}. Autoencoders have played a pivotal role in developing structured latent spaces for pose reconstruction. In this context, Milef et al.~\cite{milef2023} introduced new interpolation and pose anomaly detection methods to avoid traversing invalid regions of the latent space. Additionally, Ponton et al.~\cite{ponton2023sparseposer} combined a skeleton-aware autoencoder with learned inverse kinematics for precise full-body pose reconstruction, ensuring better accuracy in end-effectors. Other techniques~\cite{Winkler:2022, ye2022} involved the development of reinforcement learning frameworks to create natural and realistic movements. Lee et al.~\cite{lee2023} extended this to include environment interactions. These techniques have demonstrated their ability to reconstruct high-quality, smooth motion. However, they are often constrained by the rigidity of their deep-learning architectures, which limits their adaptability to varying numbers of sparse inputs or constraints. Furthermore, these methods frequently face challenges in accurately representing hard constraints, such as precise end-effector positioning.

\paragraph*{Motion priors for Motion Capture}
Leveraging compressed latent space representations has become essential for human motion synthesis due to the challenges of working directly with large amounts of mocap data. Early work~\cite{andrews:2016} used a physics-based model and a motion prior to reconstruct poses from a set of IMUs and optical markers. Subsequent work~\cite{ling:2020, peng:2022} explored VAEs for encoding and decoding poses, thereby enabling deep reinforcement learning in the latent space. In the context of learning neural representations as priors, Liu et al.~\cite{liu_learning_2022} introduced Lipschitz regularization to enforce continuity and facilitate smooth interpolation and extrapolation operations. Motion priors have proven valuable for handling noisy or faulty pose data. For instance, Rempe et al.~\cite{rempe:2021} achieved impressive results by training a motion prior on the AMASS dataset, later refined by Shi et al.~\cite{shi:2023} with a periodic autoencoder~\cite{starke:2022}. Once their latent space is trained, pose transition sequences are optimized to satisfy some given constraints, such as end-effector positions or fitting a point cloud. However, challenges remain in reproducing high-frequency details and real-time processing. Working with transitions limits the ability to make targeted adjustments to individual poses within the latent space, as each pose is dependent on the entire preceding sequence. Our approach addresses these limitations by representing full poses directly in the latent space. Direct pose optimization in the latent space allows our method to have a more fine-grained control while being performant, and ensures constraints are met frame by frame instead of optimizing large motion sequences.

\subsection{Learned Inverse Kinematics}

In computer animation, IK solvers are essential for determining the positions and orientations of intermediate joints in a kinematic chain, given the target positions and orientations of end-effectors. Aristidou et al.~\cite{Aristidou:2018} offers a comprehensive review of popular IK approaches for human motion reconstruction. Yet, traditional IK solvers often face scalability issues for multi-chain characters and a balance must be struck between computational efficiency and pose naturalness~\cite{Caserman:2019}.


Early work~\cite{Grochow:2004, Wu:2011} introduced a data-driven IK system using Gaussian processes for versatile pose modeling. The work by Huang et al.~\cite{Huang:2017} further enhanced the idea of using Gaussian models with a traditional Jacobian-based solver for real-time pose generation. With the augmented capabilities of modern deep-learning-based architectures, Victor et al.~\cite{Victor:2021} proposed an autoencoder-based IK solver. A common issue in deep learning-based IK methods is their fixed network structure, which limits their use to a specific number of sparse inputs. To address this, Oreshkin et al.~\cite{oreshkin2021a} implemented prototype encoding and residual connections (ProtoRes), which enables dynamic adjustments on the number of end-effectors. Building on this, Voleti et al.~\cite{voleti2022} adapted the method for SMPL models~\cite{loper2015} and incorporated initial pose estimation from images. Further advancements were made by Agrawal et al.~\cite{Agrawal:2023}, who refined the technique to ensure pose consistency across changes. They utilized a skeletal graph structure, enabling the encoding of hard constraints by restricting information flow at certain joints. This adaptation significantly enhances the workflow for artists by allowing them to maintain base poses while editing specific body parts. However, protores-based systems still contain predefined constraints embedded within the network structure, limiting the ability to modify these constraints dynamically. Moreover, as both constraints and end-effectors variability are encoded as inputs, the network requires training across all possible permutations of sparse input and constraints, posing scalability challenges.

\section{Background}
\label{sec:method:vae}

In this section, we provide an overview of Variational Autoencoders (VAE), the core concept behind our Pose Autoencoder design, detailed in Section~\ref{sec:method:architecture}. This design is key in creating a structured latent space, which is subsequently utilized in the Pose Optimizer step outlined in Section~\ref{sec:method:inference}.

\paragraph*{Variational Autoencoders}
Unlike traditional autoencoders, which generate a latent space with no continuity guarantee, VAEs explicitly aim to construct a latent space that is continuous, enabling the generation of data through the interpolation of latent representations, as well as generating random data by sampling randomly the latent space. More formally, given a set of continuous observations $\{ \mathbf{x}^{(i)} \}^{\scriptscriptstyle N}_{\scriptscriptstyle i=1}$ let us assume that an observation can be reconstructed from an unobserved continuous random variable $\mathbf{z}$, which follows a gaussian distribution $p(\mathbf{z}) = \mathcal{N}(0, \mathbf{I})$. Next, the objective is to find a decoder $\mathcal{D}_{\boldsymbol{\theta}}$ with parameters $\boldsymbol{\theta}$ which maximize the marginal log-likelihood as follows:
\begin{equation}
    \log p_{\boldsymbol{\theta}}(\mathbf{x}^{(i)}) = \log \int_{\mathbf{z}} p_{\boldsymbol{\theta}}(\mathbf{x}^{(i)} | \mathbf{z}) p(\mathbf{z}) \, d\mathbf{z}
\end{equation}
However, optimizing the marginal log-likelihood is not feasible due to the intractability of the integral. Therefore, the Evidence Lower Bound (Eq.~\ref{eq:elbo}) can be used to obtain a feasible lower bound of $\log p_{\boldsymbol{\theta}}(\mathbf{x}^{(i)})$. 
\begin{equation}
\label{eq:elbo}
    -D_{KL} \left( q_{\boldsymbol{\phi}}(\mathbf{z}|\mathbf{x}^{(i)}) \, || \, p(\mathbf{z}) \right) + \mathbb{E}_{\mathbf{z} \sim q_{\boldsymbol{\phi}}(\mathbf{z}|\mathbf{x}^{(i)})} \left[ \log p_{\boldsymbol{\theta}}(\mathbf{x}^{(i)} | \mathbf{z}) \right]
\end{equation}
where $q_{\boldsymbol{\phi}}(\mathbf{z}|\mathbf{x}^{(i)})$ is an approximation of the posterior $p_{\boldsymbol{\theta}}(\mathbf{z}|\mathbf{x}^{(i)})$ with the learned parameters $\boldsymbol{\phi}$. The approximation of the posterior $q_{\boldsymbol{\phi}}$ can be represented as a diagonal Gaussian, with means and variance predicted by the encoder $\mathcal{E}_{\boldsymbol{\phi}} = ( \mu_{\boldsymbol{\phi}}, \sigma_{\boldsymbol{\phi}} )$, as follows:
\begin{equation}
    q_{\boldsymbol{\phi}}(\mathbf{z}|\mathbf{x}^{(i)}) = \mathcal{N} \left( \mu_{\boldsymbol{\phi}}(\mathbf{x}^{(i)}), \mathbf{I} \, \sigma_{\boldsymbol{\phi}}(\mathbf{x}^{(i)}) \right)
\end{equation}
To optimize the parameters $( \boldsymbol{\theta} , \boldsymbol{\phi} )$ the goal is to maximize Eq.~\ref{eq:elbo}. The first term is the KL divergence, denoted as $\mathcal{L}_{KLD}$ in the following sections. This term encourages the $q_{\boldsymbol{\phi}}$ distribution to closely resemble the normal distribution $\mathcal{N}(\mathbf{0}, \mathbf{I})$. Additionally, it structures the latent space effectively, ensuring that when sampling $\mathbf{z} \sim \mathcal{N}(\mathbf{0}, \mathbf{I})$ and passing it through the decoder $\mathcal{D}_{\boldsymbol{\theta}}(\mathbf{z})$ a valid observation $\mathbf{\hat{x}}$ is generated. The second term of Eq.\ref{eq:elbo} is the autoencoder reconstruction loss. We will define it later as a combination of losses specifically crafted for our problem.

\section{Method}
\label{sec:method}

In this section, we first state the problem in Section~\ref{sec:method:prob_def}, and then we provide an overview of the proposed solution (refer to Section~\ref{sec:method:overview}). Subsequently, we present the architecture designed for full-body motion reconstruction from sparse input in Section~\ref{sec:method:architecture}, while, in Section~\ref{sec:method:inference}, we offer a detailed explanation of how these components are employed during inference.

\subsection{Problem Definition}
\label{sec:method:prob_def}
Human motion can be defined as an ordered sequence of $T$ poses $( \mathbf{x}^{(i)} )^{\scriptscriptstyle T}_{\scriptscriptstyle i=1}$. These poses are based on a humanoid skeleton comprising $J$ joints. In contrast to common hierarchical skeleton definitions, where joints are usually defined within the local frame of their parent, we adopt a distinct approach. In our model, joints are defined relative to the root space, utilizing quaternions. In order to ensure that our model is invariant to global transformations, we avoid encoding the root joint with explicit global information. Instead, the root joint is expressed through a quaternion encoding the root's rotational increment and a 3D vector encoding displacement, both with respect to the previous pose, making $\mathbf{x}^{(i)} \in \mathbb{R}^{J \times 4 + 3}$.

In our system, the input comprises an ordered sequence of $T$ sparse poses $\mathbf{s}^{(i)}$, encompassing only a subset of joints $S \leq J$ (e.g., end-effector data from VR controllers). Unlike complete poses, each sparse pose is characterized by the global positions (represented as 3D vectors) and rotations (represented as quaternions) of the sparse tracking signals, resulting in $\mathbf{s}^{(i)} \in \mathbb{R}^{S \times (3 + 4)}$. The objective of our task is to synthesize full-body human motion $( \mathbf{x}^{(i)} )^{\scriptscriptstyle T}_{\scriptscriptstyle i=1}$ from sparse inputs $( \mathbf{s}^{(i)} )^{\scriptscriptstyle T}_{\scriptscriptstyle i=1}$.

\subsection{Overview}
\label{sec:method:overview}
\begin{figure*}[ht]
  \includegraphics[width=1\linewidth]{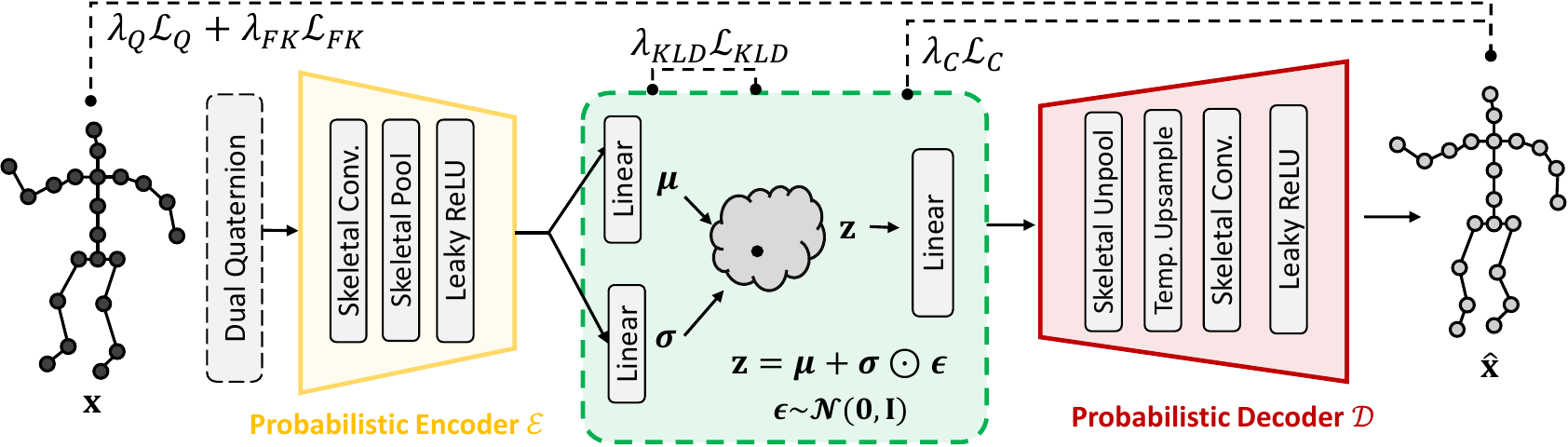}
  \caption{Pose Autoencoder architecture for creating a structured latent space representing valid poses. Following the Variational Autoencoder approach, we train the network, beginning with the input pose data represented in dual quaternions to capture joint rotations and translations in root space. The Probabilistic Encoder $\mathcal{E}$ estimates the mean and variance parameters of the Gaussian distribution for the input pose. Subsequently, a latent vector $\mathbf{z}$ is sampled and passed into the Probabilistic Decoder $\mathcal{D}$ for pose reconstruction. Training involves two reconstruction losses: $\mathcal{L}_Q$ and $\mathcal{L}_{FK}$, in addition to $\mathcal{L}_{KLD}$ and $\mathcal{L}_C$ which structure the latent space, ensuring standard Gaussian alignment and latent space continuity, respectively.}
  \label{fig:vae_pipeline}
\end{figure*}

This section outlines our method for synthesizing continuous full-body motion from sparse input. DragPoser is comprised of a Pose Autoencoder and a Temporal Predictor. These components are used in the pose optimizer step to generate the final motion, as depicted in Figure~\ref{fig:inference_pipeline}. Our pose optimizer method relies on having a structured and continuous latent space generated by the pose autoencoder (see Figure~\ref{fig:vae_pipeline}). During the optimization process, we start with an initial pose and a set of sparse tracking signals. We then conduct an optimization-based search within the latent space to identify a feasible pose that meets our specified constraints. This direct optimization in the latent space enables us to accurately fulfill hard constraints and generate human-like poses similar to those the pose autoencoder was trained on. Additionally, a temporal predictor guides the optimization process to ensure temporal consistency between poses.

\subsection{Architecture}
\label{sec:method:architecture}

\paragraph*{Pose Autoencoder}
Our motion synthesis method relies on an optimization process in a latent space (explained in Section~\ref{sec:method:inference}). The Pose Autoencoder component is used to construct this latent space from motion capture data. Autoencoders rely on the idea of having a bottleneck between two networks\textemdash Encoder and Decoder\textemdash which allows the networks to effectively project the supplied data to a space with fewer dimensions, known as latent space. However, this latent space is not explicitly structured and, thus, may have regions with invalid points. To address this, we based our Pose Autoencoder on the Variational Autoencoder (VAE) architecture (see Section~\ref{sec:method:vae}), which forces the latent space to represent valid points in a specific domain. To design the Encoder and Decoder networks, we require a network capable of accurately reconstructing a large amount of motion capture data. We build on top of the work by Ponton et al.~\cite{ponton2023sparseposer} which proposes an autoencoder structure, with no explicit structure on the latent space, that uses skeleton-aware operations as building blocks \cite{Aberman:2020}. Next, we introduce all modifications needed to construct our Pose Autoencoder; please refer to Ponton et al.~\cite{ponton2023sparseposer} for details on the skeleton-aware operations.

\paragraph*{Probabilistic Encoder}
Our Pose Autoencoder reconstructs single poses $\mathbf{x}^{(i)} \in \mathbb{R}^{J \times 4 + 3}$ as detailed in Section~\ref{sec:method:prob_def}. However, the input of the Encoder $\mathcal{E}$ is extended with each joint's root space translations. This information helps the Encoder understand human motion better by explicitly encoding information about the dimensions of the skeleton. For this, we use dual quaternions~\cite{Andreou:2022}, which provide a unified and compact representation encoding both the rotational and translation information in orthogonal quaternions, as used in previous work~\cite{ponton2023sparseposer}. We encode the root's world displacement in its corresponding dual quaternion. Thus, we define the function $DQ : \mathbb{R}^{J \times 4 + 3} \mapsto \mathbb{R}^{J \times 8}$, converting our original representation to dual quaternions. 

Given a pose $\mathbf{x}^{(i)}$, we convert it to the dual quaternions representation $DQ(\mathbf{x}^{(i)})$ and use the Encoder to project it to a latent vector $\mathbf{z}^{(i)} \in \mathbb{R}^L$ with dimension $L$ (in our tests $L=24$ for $J=22$). However, as explained in Section~\ref{sec:method:vae}, we do not use the Encoder's output directly as the latent vector. We use two linear layers to project the Encoder's output to the mean and variance with $L$ dimensions. Thus, we represent the Encoder $\mathcal{E}_{\boldsymbol{\phi}}$ with parameters $\boldsymbol{\phi}$ as the tuple $( \mu_{\boldsymbol{\phi}}, \sigma_{\boldsymbol{\phi}} )$. The final mean and variance represent the Gaussian distribution of the input observation in the latent space.
The resulting latent vector $\mathbf{z}^{(i)}$ is computed as follows:
\begin{equation}
    \mathbf{z}^{(i)} = \mu_{\boldsymbol{\phi}} \! \left( DQ(\mathbf{x}^{(i)})\right) + \mathbf{I} \, \sigma_{\boldsymbol{\phi}} \! \left(DQ(\mathbf{x}^{(i)})\right) \epsilon
\end{equation}
where $\epsilon \sim \mathcal{N}(0, \mathbf{I})$.

\paragraph*{Probabilistic Decoder}
Once the latent vector $\mathbf{z}^{(i)}$ is formed, we use a linear layer and the Decoder $\mathcal{D}_{\boldsymbol{\theta}}$ with parameters $\boldsymbol{\theta}$ to reconstruct the original pose $\mathbf{x}^{(i)}$ as follows:
\begin{equation}
    \mathbf{\hat{x}}^{(i)} = \mathcal{D}_{\boldsymbol{\theta}}(\mathbf{z}^{(i)})
\end{equation}

\paragraph*{Reconstruction Loss}
We utilize a regular Mean Squared Error reconstruction loss, defined as follows:
\begin{equation}
\mathcal{L}_Q = MSE(\mathbf{x}^{(i)}, \mathbf{\hat{x}}^{(i)})
\end{equation}

\paragraph*{Forward Kinematics Loss}
To facilitate the neural network's understanding of the joint hierarchy, we additionally include an FK-based loss \cite{Pavllo:2018, Pavllo:2020}. We compute the global position of each joint using FK and compare it with the ground truth. Thus, we define the function $FK: \mathbb{R}^{J\times4+3} \mapsto \mathbb{R}^{J\times3}$, and the loss as follows:
\begin{equation}
\mathcal{L}_{FK} = MSE\left(FK(\mathbf{x}^{(i)}), FK(\mathbf{\hat{x}}^{(i)})\right)
\end{equation}

\paragraph*{KL Divergence Loss}
This loss is inherent to the Variational Autoencoder architecture as introduced in Section~\ref{sec:method:vae}. It tries to force all the Encoder output distributions to have a small KL divergence regarding the desired distribution of the latent space\textemdash a Gaussian with mean zero and variance one. This allows us to traverse the latent space and find valid poses around the desired distribution in the optimization process. We define $\mathcal{L}_{KLD}$ as follows:
\begin{align}
\mathcal{L}_{KLD} &= \frac{1}{2} \langle{ 1 + \log\left((\sigma^{(i)})^2\right) -  (\mu^{(i)})^2 - (\sigma^{(i)})^2, \mathbf{1}}\rangle \\
\mu^{(i)} &= \mu_{\boldsymbol{\phi}} \! \left(DQ(\mathbf{x}^{(i)})\right) \\
\sigma^{(i)} &= \sigma_{\boldsymbol{\phi}} \! \left(DQ(\mathbf{x}^{(i)})\right)
\end{align}

\begin{figure}[htb]
  \includegraphics[width=0.75\linewidth]{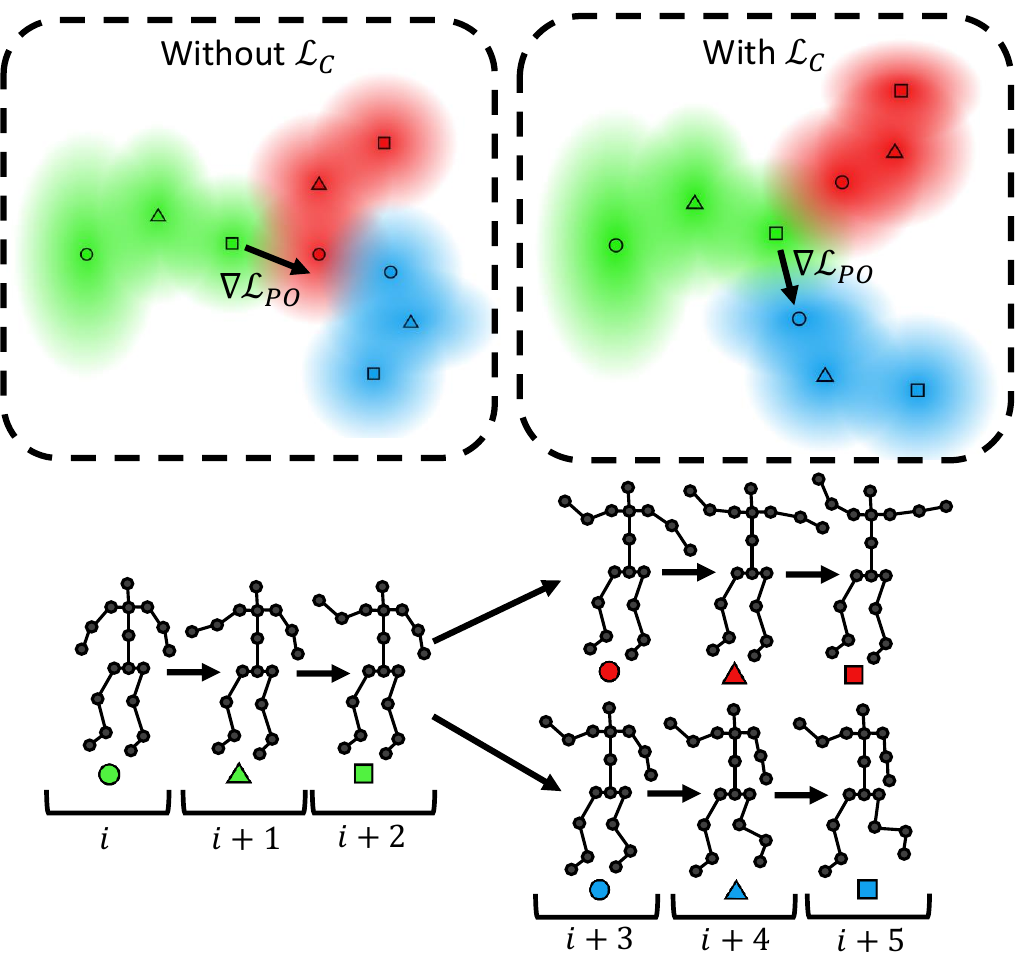}
\caption{Demonstration of how the continuity loss $\mathcal{L}_C$ structures the latent space to reflect the temporality of data. The left diagram shows a latent space learned without $\mathcal{L}_C$, where similar poses are clustered together, but transitioning smoothly from one pose to the next is not guaranteed\textemdash potentially resulting in getting stuck in local minima. The right diagram illustrates the latent space with $\mathcal{L}_C$, encouraging consecutive poses from the original motion sequence to be within reach of a single optimization step. Notice the ability to transition from the last green pose to either the first red or blue pose, representing two potential future sequences. Each pose is represented as a 2D Gaussian distribution, while the mean predicted by the Encoder is represented with a symbol.}
\label{fig:continuity_loss}
\end{figure}

\paragraph*{Continuity Loss}
Finally, we propose the Continuity Loss to improve the Pose Optimizer process (see Section~\ref{sec:method:inference}). Given two consecutive poses ($\mathbf{x}^{(i)}$, $\mathbf{x}^{(i+1)}$), we want them together in the latent space so that with one optimization step we can change from the initial pose to the next one: $\hat{\mathbf{z}}^{(i+1)} = \mathbf{z}^{(i)} - \nabla_\mathbf{z} MSE(\hat{\mathbf{x}}^{(i)}, \mathbf{x}^{(i+1)})$. 
Although the KLD Loss $\mathcal{L}_{KLD}$ enforces similar poses to be together in the latent space, it does not have the notion of temporal continuity. With the continuity loss, we further guide the latent space construction process to facilitate the Pose Optimizer process. The loss is defined as follows:
\begin{align}
&\mathcal{L}_{C} = MSE\left(\mathbf{x}^{(i+1)}, \mathcal{D}_{\boldsymbol{\theta}}(\hat{\mathbf{z}}^{(i+1)})\right) \\
& \hat{\mathbf{z}}^{(i+1)} = \mathbf{z}^{(i)} - \nabla_\mathbf{z} MSE(\hat{\mathbf{x}}^{(i)}, \mathbf{x}^{(i+1)}) \\
&\nabla_\mathbf{z} MSE(\hat{\mathbf{x}}^{(i)}, \mathbf{x}^{(i+1)}) = \mathbf{J}^\mathrm{T}_{\mathbf{z}} \mathcal{D}_{\boldsymbol{\theta}}(\mathbf{z}^{(i)}) \cdot 2(\mathbf{x}^{(i+1)} - \hat{\mathbf{x}}^{(i)})
\end{align}

where $\mathbf{J}^\mathrm{T}_{\mathbf{z}}$ is the transpose of the Jacobian matrix of the Decoder $\mathcal{D}$ with respect to $\mathbf{z}$. Figure~\ref{fig:continuity_loss} illustrates the role of the Continuity Loss ($\mathcal{L}_{C}$) in structuring the latent space based on the temporal sequence of poses. In the absence of $\mathcal{L}_{C}$, the Pose Autoencoder clusters similar poses without considering their temporal sequence, which may result in an undesirable pose being traversed between two consecutive poses during optimization. This can cause the optimization process to become trapped in local minima. Conversely, incorporating $\mathcal{L}_{C}$ encourages the latent space to organize by pose similarity and temporal succession, ensuring that one pose leads smoothly to the next. By mitigating the risk of local minima, this structure allows for more reliable pose transitions and faster optimization (demonstrated by the reduced iteration count in Section~\ref{sec:ablation}).

\paragraph*{Temporal Predictor}
\label{sec:method:temporal}

The Pose Autoencoder effectively constructs a structured latent space for decoding single poses. However, this space lacks two critical properties. First, it is unbounded, assuming arbitrary shapes as combinations of Gaussian distributions. Second, while the Continuity Loss promotes poses to be reachable during the optimization step, it remains unclear which pose to select for temporal coherence among all possible reachable poses. This implies a lack of a clear understanding of the sequential progression of poses during optimization. We explored the idea of directly encoding sequences of poses in the latent vector. However, this resulted in a very slow optimization process and mostly non-accurate results due to the sparsity of the data. To address these issues, we propose training a neural network\textemdash Temporal Predictor\textemdash to predict the most likely sequences of latent vectors based on their coherence with previous sequences. These predictions can then inform and guide the Pose Optimizer process.
This approach not only helps navigate away from invalid regions in the latent space but also provides the system with an understanding of the temporal sequence of the data. A visual overview of the Temporal Predictor is depicted in Figure~\ref{fig:inference_pipeline}.

To maximize the number of previous and future latent vector sequences, the Temporal Predictor $\mathcal{T}$ operates at a reduced frequency. Specifically, $\mathcal{T}$ is executed every $n$ frames, with $n=4$ in our experiments. We previously used $i$ to denote frames; now, we introduce a new index $j = \lceil i / n \rceil$. For indexing standard latent vectors $\mathbf{z}^{(i)}$, our frame rate is set to 60\,fps, which corresponds to the standard frame rate of the Pose Optimizer. For indexing predicted latent vectors by the Temporal Predictor $\mathbf{z_t}^{(j)}$, we use the lower frame rate. The Temporal Predictor $\mathcal{T}_{\boldsymbol{\psi}}$, with parameters $\boldsymbol{\psi}$, adopts the Transformer architecture, comprising Encoder ($\mathcal{T}^E_{\boldsymbol{\psi}}$) and Decoder ($\mathcal{T}^D_{\boldsymbol{\psi}}$) models.

The process begins with a current Temporal Predictor index $j$, a desired past window $W_p$, and a future window $W_f$ for predicted latent vectors. We define $p$ as the index for initiating new latent vector predictions. Specifically, the Encoder $\mathcal{T}^E_{\boldsymbol{\psi}}$ is activated every $W_f$ steps, with $p = W_f \lfloor j / W_f \rfloor$. Its input is a sequence of past latent vectors $( \mathbf{z}^{(n \cdot k)} ){\scriptscriptstyle p-W_p \leq k < p}$. Subsequently, the Decoder $\mathcal{T}^D_{\boldsymbol{\psi}}$, using the Encoder's output and the initial latent vector $\mathbf{z}^{(n \cdot p)}$, predicts the next latent vector $\hat{\mathbf{z}}^{(j)}$. This output is then concatenated to the input of the Decoder, and the process repeats $W_f$ times, similarly to the execution of Transformers in Natural Language Processing tasks. The Decoder's input is formally defined as $\left( \mathbf{z}^{(n \cdot p)}, ( \mathbf{z_t}^{(k)} ){\scriptscriptstyle p < k < j} \right)$. Additionally, root displacements and the height to the ground for the root, head, hands, and feet are provided to the Encoder $\mathcal{T}^E_{\boldsymbol{\psi}}$ to narrow down the search space.

The Encoder's role is contextualizing the Temporal Predictor's latent vector prediction. The iterative prediction of latent vectors in the Decoder, however, addresses different scenarios. For sparse inputs with minimal ambiguity, like six sensors (one per limb and the root), we set $W_f = 1$. Therefore, the Temporal Predictor only helps the Pose Optimizer process to stay within the distribution of valid poses. In cases of higher ambiguity, like four sensors (head, hands, and root), we increase $W_f$ to 16 or 60, letting the Decoder predict the next possible latent vectors. The Encoder handles actual past data, and the Decoder makes future predictions. Therefore, increasing $W_f$ effectively helps during the prediction process.

The Temporal Predictor $\mathcal{T}$ is trained before the optimization process to learn about the data's temporality. It is trained with a standard MSE loss function. Giving a motion clip from our database, we feed a sequence of poses to the Encoder $\mathcal{T}^E_{\boldsymbol{\psi}}$ and let the Decoder $\mathcal{T}^D_{\boldsymbol{\psi}}$ predict the following poses.

\subsection{Pose Optimizer}
\label{sec:method:inference}

\begin{figure*}[ht]
  \includegraphics[width=1\linewidth]{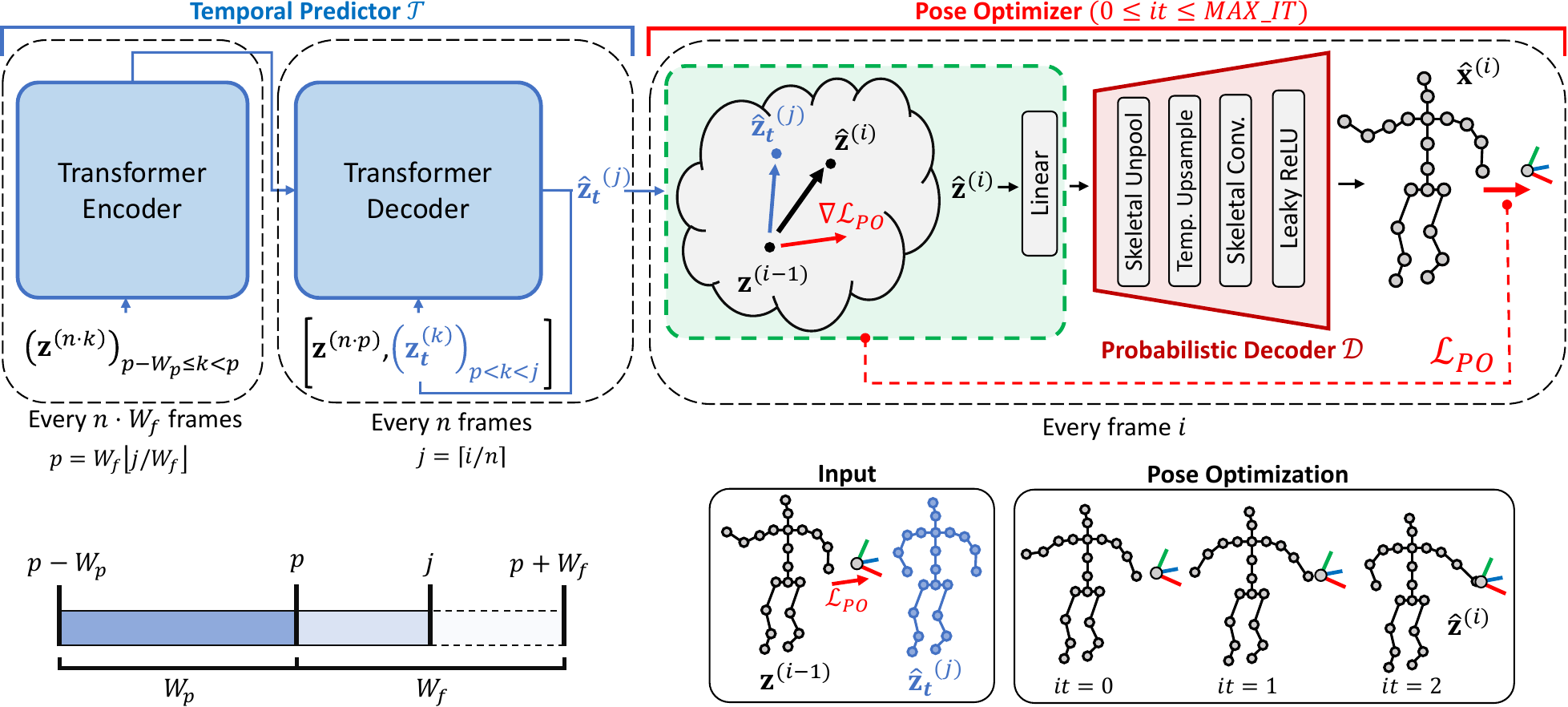}
  \caption{Motion reconstruction process via latent space optimization. The procedure begins with an initial latent vector from which the pose is decoded. Constraints are then dynamically established as loss functions, and the gradients of these losses are computed via backpropagation within the latent space. The decoded pose is iteratively refined through this process, either until a maximum iteration count is reached or certain predefined metrics are satisfied. This optimization is guided by a Temporal Predictor network that employs a Transformer architecture. The Encoder module generates features from past poses, and the Decoder module predicts future poses. While the Pose Optimizer operates at every application frame, the Temporal Predictor is executed at every $n$ frames (with $n=4$ in our experiments).}  
  \label{fig:inference_pipeline}
\end{figure*}

The preceding sections detailed the components for creating a structured latent space to generate continuous and valid poses. However, using the Pose Autoencoder, we cannot directly synthesize full-body motion from sparse input. Taking inspiration from Pan et al.~\cite{pan:2023}, who demonstrate image modification in a GAN's latent space through optimization, we crafted a similar latent space optimization process for pose prediction. This process is visually summarized in Figure~\ref{fig:inference_pipeline} and outlined in Algorithm~\ref{alg:inference}.

\begin{algorithm}[t]
\DontPrintSemicolon 
\For{$i=0$ \emph{\KwTo} $T$}{
$\hat{\mathbf{z}}^{(i)} \leftarrow \mathbf{z}^{(i-1)}$\; 
$\mathbf{s}^{(i)} \leftarrow \text{Target Sparse Input}$\;
$j \leftarrow \lceil i/n \rceil$\; 
$p \leftarrow W_f\lfloor j/W_f \rfloor$\;
\lIf{$j$ \% $W_f$ == $0$}{$\mathbf{h} \leftarrow  \mathcal{T}^E_{\boldsymbol{\psi}}\left(( \mathbf{z}^{(n \cdot k)} ){\scriptscriptstyle p-W_p \leq k < p}\right)$}
\lIf{$i$ \% $n$ == $0$}{$\mathbf{z_t}^{(j)} \leftarrow \mathcal{T}^D_{\boldsymbol{\psi}}\left( \mathbf{z}^{(n \cdot p)}, ( \mathbf{z_t}^{(k)} ){\scriptscriptstyle p < k < j}; \mathbf{h} \right)$}
\While{$MAX\_IT$ and $\mathcal{L}_{PO} > \epsilon$}
{
    $\hat{\mathbf{x}}^{(i)} \leftarrow \mathcal{D}_{\theta}(\hat{\mathbf{z}}^{(i)})$ \;

    $\mathcal{L}_{PO} \leftarrow MSE\left(FK'(\mathbf{\hat{x}}^{(i)}), \mathbf{s}^{(i)}\right)$
    
    $\hat{\mathbf{z}}^{(i)} \leftarrow \hat{\mathbf{z}}^{(i)} - \lambda_{PO}\nabla_{\mathbf{z}}\mathcal{L}_{PO} - \lambda_T\nabla_{\mathbf{z}} MSE(\hat{\mathbf{z}}^{(i)}, \mathbf{z_t}^{(j)})$\;
}
$\mathbf{z}^{(i)} \leftarrow \hat{\mathbf{z}}^{(i)}$\;
}

\caption{Motion synthesis from sparse input $( \mathbf{s}^{(i)} )^{\scriptscriptstyle T}_{\scriptscriptstyle i=1}$.}
\label{alg:inference}
\end{algorithm}

Given a frame $i$, with its sparse input $\mathbf{s}^{(i)}$, and an initial pose in latent form $\mathbf{z}^{(i-1)}$, our goal is to predict the pose $\hat{\mathbf{x}}^{(i)}$ that best satisfies the constraints defined. We begin by decoding the initial latent vector to a pose using the Decoder of the Pose Autoencoder with fixed parameters: $\hat{\mathbf{x}}^{(i-1)} = \mathcal{D}_{\boldsymbol{\theta}}(\mathbf{z}^{(i-1)})$. We then employ the sparse Forward Kinematics function $FK': \mathbb{R}^{J\times4+3} \mapsto \mathbb{R}^{S\times(4+3)}$ to match the full-body pose with the global joint positions and rotations of the sparse input. The pose optimization loss, $\mathcal{L}_{PO} = MSE\left(FK'(\hat{\mathbf{x}}^{(i)}), \mathbf{s}^{(i)}\right)$, encapsulates our constraints, which can be extended as needed, for instance, by adding or removing sensors on the fly, or imposing arbitrary limits on specific joints, or constraining only positions or rotations.

We backpropagate this loss through the Decoder to the latent space, obtaining a gradient to guide the update of $\mathbf{z}^{(i-1)}$ towards fulfilling the constraints. This process is similar to that of training the Decoder, but this time the parameters are fixed. The updated latent vector, $\hat{\mathbf{z}}^{(i)} = \mathbf{z}^{(i-1)} - \lambda_{PO} \nabla_z \mathcal{L}_{PO}$, is then decoded back to a pose $\hat{\mathbf{x}}^{(i)} = \mathcal{D}_{\boldsymbol{\theta}}(\hat{\mathbf{z}}^{(i)})$. This iterative process is repeated until the positional and rotational error thresholds of the sparse tracking signals are satisfied, a maximum iteration count is reached, or the gradient of the loss is very close to zero (e.g., $1 \times 10^{-5}$). We found the latter condition to significantly speed up computations, as it is unlikely that any improvements will be made in this case.

Additionally, we integrate the Temporal Predictor's output (see Section~\ref{sec:method:temporal}), $\mathbf{z_t}^{(j)}$, to avoid invalid latent space regions and to maintain temporal coherence, particularly when sensor input is sparse. This dual optimization, combining pose and temporal constraints, ensures our predictions remain both valid within the latent space and temporally consistent with the input data. The final latent vector update, as shown in Algorithm~\ref{alg:inference}, incorporates both constraints. We maintain $\lambda_{PO}=1$ and adaptively adjust $\lambda_T$ based on the number of sensors used (0.15, 0.125, 0.02 for 3, 4, and 6 sensors, respectively).

\section{Experiments and Evaluation}
In this section, we comprehensively evaluate DragPoser, employing both quantitative and qualitative assessments on publicly available datasets, and comparing its performance with state-of-the-art methods. Furthermore, we evaluate our system across various sensor setups and constraint configurations. Finally, we conduct an ablation study to examine the impact of DragPoser's core components. Unless otherwise noted, our evaluations utilize 6 degrees of freedom (6-DoF) signals, which provide both positional and rotational information. Default end-effector placements are: hip, head, hands, and feet for 6 signals; hip, head, hands, and right foot for 5 signals; hip, head, and hands for 4 signals; head and hands for 3 signals.

\paragraph*{Datasets}
For evaluation, we trained DragPoser using the DanceDB dataset \cite{DanceDB:2019} and evaluated with the HUMAND4D \cite{HUMAN4D:2020} and SOMA \cite{SOMA:2021} datasets, both part of the AMASS collection \cite{AMASS:2019}. Our choice of AMASS was guided by its public availability and its use in previous works. Additionally, for some images and supplemental material, we employed the SparsePoser database \cite{ponton2023sparseposer}. This database is particularly advantageous for VR applications, offering diverse common motions and reducing incorrect poses through the use of IMU-based motion capture compared to the optical systems used in AMASS datasets. The Pose Optimizer process is initialized with the first pose of each motion clip.

\paragraph*{Metrics}
To evaluate the effectiveness of our method, we employ four distinct metrics while ensuring the root position is aligned with the ground truth data for accurate pose evaluation. First, we measure the overall pose quality through three metrics: the \emph{Positional Error} (Pos), which calculates the mean Euclidean distance between corresponding joints in centimeters; the \emph{Rotational Error} (Rot), quantifying the mean angular difference between rotations $R_0$ and $R_1$ using the angle of the rotation matrix $D = R_0 R_1^T$ and the \emph{Velocity Error} (Vel), which calculates the mean velocity error across all joints in centimeters per second. Additionally, we evaluate hard constraints by computing the mean Euclidean distance error of the provided end-effectors, excluding the root, in centimeters.

\paragraph*{Training and Implementation Details}
We implemented DragPoser using PyTorch and optimized the networks with the AdamW optimizer~\cite{Loshchilov:2019}. The system was trained and evaluated on a PC configured with an Intel Core i7-12700k CPU, 32GB of RAM, and an NVIDIA GeForce RTX 3090 GPU. The Pose Autoencoder was trained with a batch size of 64 and a learning rate of $10^{-4}$ over approximately one day; we set the latent space dimension to $L=24$ and the number of joints to $J=22$, resulting in about 168 thousand parameters. For the Temporal Predictor, we trained it with a batch size of 512 and a learning rate of $10^{-3}$ for approximately thirty minutes. The Transformer hyperparameters were set to four heads, three layers each for the encoder and decoder, a feedforward dimension of 2048, and a feature dimension of 48, totaling approximately 1.2 million parameters. During training, we used $W_p = 16$ and $W_f = 16$ and ran the Pose Autoencoder's Encoder to generate the training database for the Temporal Predictor. Additionally, we empirically found that introducing random noise to the four limbs aids in regularizing the Temporal Predictor. Thus, in every training iteration, we introduce Gaussian noise to the latent vector, with mean and standard deviation values for each dimension derived from the dataset, applied to each limb with a 10\% probability. 
We initially experimented with adding noise to arbitrary joints in the full skeleton, but we found that targeting individual limbs was more effective.
First we train the Pose Autoencoder with a weighted combination of all the losses presented $\mathcal{L} = \lambda_Q \mathcal{L}_Q + \lambda_{FK} \mathcal{L}_{FK} + \lambda_{KLD} \mathcal{L}_{KLD} + \lambda_C \mathcal{L}_C$ where $\lambda_Q = 1$, $\lambda_{FK} = 100$, $\lambda_{KLD} = 0.001$ and $\lambda_C = 1$. Subsequently, we train the Temporal Predictor with a standard MSE loss.

\subsection{Comparison}
We compare DragPoser with three state-of-the-art deep-learning-based methods, AvatarPoser~\cite{Jiang:2022b}, SparsePoser~\cite{ponton2023sparseposer} and HuMoR~\cite{rempe:2021}, as well as FinalIK, a state-of-the-art IK method for animating full-body characters. These methods were selected due to their public availability and their representation of different sensor configurations and approaches.

AvatarPoser utilizes a Transformer encoder to generate full-body poses from sparse input (three or four sensors) and refines the arms with an optimization-based IK. Notably, it excels in producing high-quality poses in scenarios with high ambiguity and limited sensors. We extended the method, referred to here as AvatarPoser+, by adding two additional sensors on the feet, allowing a fair comparison in six sensor setups. Note that AvatarPoser was retrained for each different experiment due to its fixed input.

In addition, we compare with SparsePoser, the state-of-the-art deep-learning-based method in terms of pose quality and end-effector accuracy for six 6-DoF sensors. Their approach is fixed to six input sensors and uses an autoencoder and learned IK networks for full-body pose reconstruction.

We also compare it with HuMoR, which, similar to our work, uses a VAE autoencoder to construct a latent space later used during an optimization process for multiple tasks. Conversely, HuMoR employs a VAE autoencoder on pose transitions rather than complete poses. Additionally, the provided implementation requires processing entire motion sequences, increasing computational demands and leaving the method unsuitable for real-time or interactive applications. Moreover, motion sequence length in HuMoR is severely constrained by available memory, requiring batch processing and additional techniques to maintain continuity. For this reason, we executed HuMoR in batches of 60 frames as recommended in the original paper since we experimented with longer batches, but the resulting animations were often highly inaccurate.

AvatarPoser, SparsePoser and HuMoR were re-trained using the DanceDB dataset following the training procedure of their official implementations. Finally, considering the findings by Ponton et al.~\cite{ponton2023sparseposer} that FinalIK exhibits superior end-effector accuracy compared to deep learning-based methods, it is employed as a baseline for evaluating hard constraints.

\begin{table}
    \caption{Accuracy of the reconstructed poses compared against state-of-the-art methods, for diverse sensor configurations.  We report the mean and standard deviation (within parenthesis) of joint errors (position, rotation, and velocity) as well as the end-effector positional error, all of them with respect to ground truth data. Please refer to the text for details about training and test datasets. When using 6 sensors (placed at the hip, head, hands, and feet), we compare against AvatarPoser+ (i.e., \cite{Jiang:2022} extended to use also the two sensors on the feet), FinalIK, HuMoR \cite{rempe:2021} and SparsePoser \cite{ponton2023sparseposer}. For the 4-sensor configuration (hip, head, and hands), we compare against AvatarPoser and HuMoR; FinalIK is not considered since it requires all end-effectors, and SparsePoser's architecture is designed for a fixed 6-sensor configuration. Concerning the 3-sensor setup, we tested our approach with the head+hands and hip+hands configurations, and compared it against the default head+hands AvatarPoser configuration and HuMoR.   
    \\
    \textsuperscript{\textdagger}HuMoR reconstructs motion in sequences, allowing it to leverage future information to adjust past poses and, thus, refine entire motion sequences during offline processing\textemdash in contrast to the other real-time, frame-to-frame prediction methods used in the comparison.}
    \scriptsize
    \centering
    \begin{tabular}{lccccc}
        \toprule
         \multirow{3}{*}{Method} & \multirow{3}{*}{Sensors} & \multicolumn{1}{c}{End-Effect. Err.} & \multicolumn{3}{c}{Joint Error} \\
         \cmidrule(lr){4-6} \cmidrule(lr){3-3}
         & & Pos (\textit{cm}) & Rot (\textit{deg}) & Vel (\textit{cm/s}) & Pos (\textit{cm}) \\
        \midrule
        AvatarPoser+ & 6 & 11.2(7.93) & 11.3(10.5) & 13.4(28.5) & 7.66(7.60) \\
        FinalIK & 6 & 1.16(1.22) & 11.5(19.2) & \textbf{6.18(21.2)} & 3.31(4.64) \\
        HuMoR\textsuperscript{\textdagger} & 6 & 9.25(4.91) & 9.35(10.4) & 19.8(35.9) & 6.96(5.27) \\
        SparsePoser & 6 & 3.43(2.53) & \textbf{5.78(5.85)} & 10.8(15.0) & 2.81(2.66) \\
        Ours & 6 & \textbf{1.02(0.73)} & 7.66(8.71) & 10.7(15.1) & \textbf{2.18(2.70)} \\
        \midrule
        AvatarPoser & 4 & 9.62(6.51) & 12.4(11.7) & \textbf{16.3(31.7)} & 9.24(10.1) \\
        HuMoR\textsuperscript{\textdagger} & 4 & 8.61(4.14) & 10.8(11.3) & 25.6(56.4) & 10.9(11.6) \\
        Ours & 4 & \textbf{1.12(0.72)} & \textbf{10.6(12.4)} & 20.0(41.5) & \textbf{7.86(13.1)} \\
        \midrule
        AvatarPoser & \multirow{2}{*}{3} & \multirow{2}{*}{12.3(7.51)} & \multirow{2}{*}{14.3(12.8)} & \multirow{2}{*}{\textbf{20.1(39.5)}} & \multirow{2}{*}{11.0(10.8)} \\
        \vspace{0.5mm}
        \emph{\tiny{Head+Hands}} & & & & & \\
        HuMoR\textsuperscript{\textdagger} & \multirow{2}{*}{3} & \multirow{2}{*}{15.0(8.30)} & \multirow{2}{*}{\textbf{11.3(12.5)}} & \multirow{2}{*}{28.0(65.1)} & \multirow{2}{*}{14.2(13.0)} \\ 
        \vspace{0.5mm}
        \emph{\tiny{Head+Hands}} & & & & & \\
        HuMoR\textsuperscript{\textdagger} & \multirow{2}{*}{3} & \multirow{2}{*}{9.92(4.71)} & \multirow{2}{*}{11.6(12.6)} & \multirow{2}{*}{27.6(63.0)} & \multirow{2}{*}{13.4(12.8)} \\
        \vspace{0.5mm}
        \emph{\tiny{Hip+Hands}} & & & & & \\
        Ours & \multirow{2}{*}{3} & \multirow{2}{*}{16.1(11.4)} & \multirow{2}{*}{22.0(22.3)} & \multirow{2}{*}{34.0(50.2)} & \multirow{2}{*}{15.5(15.0)} \\
        \vspace{0.5mm}
        \emph{\tiny{Head+Hands}} & & & & & \\
        Ours & \multirow{2}{*}{3} & \multirow{2}{*}{\textbf{1.77(1.31)}} & \multirow{2}{*}{14.1(15.6)} & \multirow{2}{*}{23.1(36.2)} & \multirow{2}{*}{\textbf{9.95(13.7)}} \\
        \vspace{0.5mm}
        \emph{\tiny{Hip+Hands}} & & & & & \\
        \bottomrule
    \end{tabular}
    \label{tab:eval:comparison}
\end{table}

\paragraph*{Quantitative}
Table~\ref{tab:eval:comparison} provides an overview of quantitative comparisons, highlighting DragPoser's effective balance between IK-like hard constraint representation and high-quality pose reconstruction, an attribute commonly associated with deep-learning methods. In these experiments, we found that DragPoser faces limitations in high-ambiguity three-sensor scenarios, particularly when lacking hip sensor data, resulting in performance below that of AvatarPoser.

In the six sensors setup (head, hands, hip, and feet), DragPoser surpasses AvatarPoser+, FinalIK and HuMoR in pose quality, aligning with SparsePoser's performance in \emph{Joint Error}. Most importantly, DragPoser excels in \emph{End-Effector Error}, outperforming all deep-learning methods and slightly improving results compared to FinalIK. This superiority is primarily attributed to the Pose Optimization process in the latent space, which utilizes the Pose Autoencoder's Decoder for achieving deep-learning-like pose quality, while concurrently allowing for the fulfillment of hard constraints as traditional IK. Despite the enhancement in \emph{Positional Error}, DragPoser encounters a slight increase in \emph{Rotational Error}. Finally, we observe that the \emph{Velocity Error} performance mirrors that of SparsePoser, with FinalIK having the best performance but with a higher standard deviation.

We also experimented with further reducing the input signals to four sensors placed in the head, hands and hip. This is a highly challenging scenario for leg pose reconstruction. In this context, DragPoser exhibits a slight advantage over AvatarPoser and HuMoR in positional and rotational errors, however, with a slightly higher \emph{Velocity Error} compared to AvatarPoser. It is important to note that, despite the reduced sensor count, DragPoser continues to maintain high end-effector accuracy (in this case, we take into account only the four sensors). 

However, when only three sensors are used, and the hip sensor is removed, DragPoser struggles to achieve the pose quality seen in AvatarPoser and HuMoR. On the other hand, when the three sensors are placed on the hip and hands, DragPoser attains comparable results to AvatarPoser and HuMoR when robust constraints are incorporated. We attribute this limitation in our method to the absence of a dedicated component for global position prediction in the case of AvatarPoser, and not optimizing at a sequence level such as HuMoR. However, we decided to prioritize frame-to-frame prediction to use our method in an online prediction setting. As a result, when no hip sensor is available, the method optimizes the outcome solely based on the provided sensors and the Temporal Predictor, which proved insufficient for achieving accurate global positioning. To address this, we dynamically introduce losses, such as enforcing proximity of the feet to the floor and minimizing the distance between hip and head ground projections. While this adaptability highlights DragPoser's capability to swiftly integrate new constraints, it also emphasizes the ongoing challenge of achieving precise global position accuracy without a hip sensor.

In summary, our findings demonstrate that DragPoser achieves state-of-the-art performance across diverse end-effector scenarios and excels in pose reconstruction, particularly when high end-effector accuracy is crucial. In addition, DragPoser proves to be a flexible method concerning input configuration and robustness to occlusions and sensor malfunctions, as indicated in Table~\ref{tab:eval:sensors}.

\begin{figure}[ht]
  \includegraphics[width=0.95\linewidth]{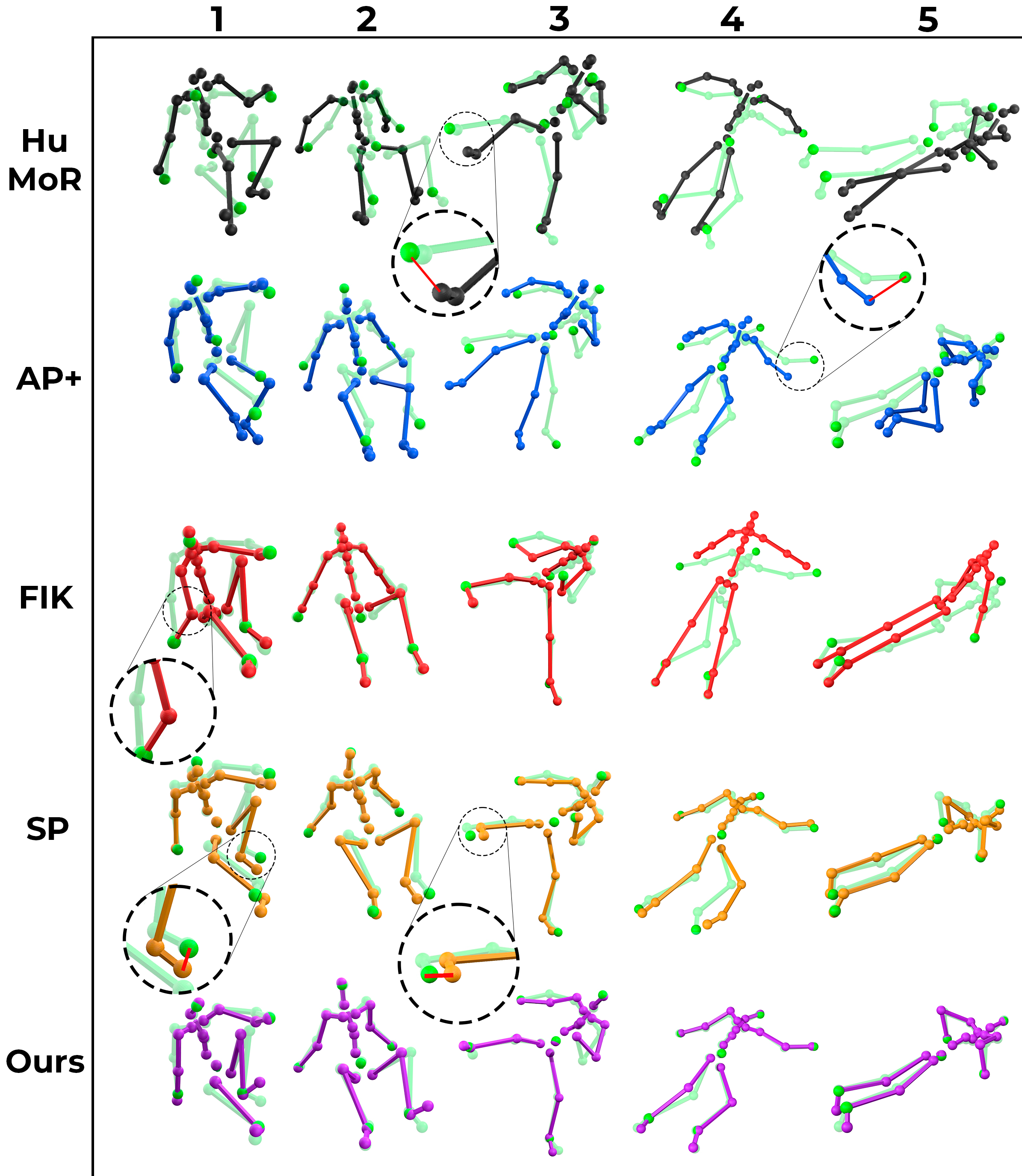}
  \caption{Visual comparison of the pose reconstruction results from HuMoR, AvatarPoser+ (AP+), FinalIK (FIK), SparsePoser (SP), and DragPoser (Ours). The ground truth poses are highlighted in green and overlaid on the results. All methods are reconstructed using six 6-DOF trackers as input data. We highlighted with a zoom-in some examples of errors found when compared against the ground truth. Deep-learning-based methods (HuMoR, AvatarPoser+ and SparsePoser) often cannot accurately place end-effectors, while traditional IK (FinalIK) produces unnatural poses. Our method can produce high-quality poses while achieving high end-effector accuracy.}
  \label{fig:methods_comparison}
\end{figure}

\paragraph*{Qualitative}
We present a visual comparison in Figure~\ref{fig:methods_comparison}, showcasing selected frames from different poses in the evaluation datasets. In the text, we refer to specific poses in Figure~\ref{fig:methods_comparison} as row/column. All methods utilize six 6-DoF sensors, with end-effectors represented by green spheres. Note that some methods\textemdash HuMoR, FinalIK and DragPoser\textemdash use the ankle joints for foot end-effectors, while the others\textemdash AvatarPoser+ and SparsePoser\textemdash use the toe joints.

Generally, FinalIK demonstrates accurate end-effector placement but often at the cost of losing plausible poses. \emph{FIK/1} and \emph{FIK/3} exemplify this, with incorrectly oriented elbows and significant deviations from ground truth in the rest of the examples. Moreover, due to its reliance on ad-hoc solutions for natural pose generation, FinalIK struggles in less common scenarios, failing to reconstruct poses or reach end-effectors, as seen in the \emph{FIK/4} and \emph{FIK/5}.

HuMoR, AvatarPoser+ and SparsePoser encounter similar challenges. While they are able to decode high-quality poses, there are instances, such as in the \emph{AP+/3}, \emph{AP+/5} or \emph{HuMoR/5}, where they fail to accurately reconstruct poses. In terms of end-effectors, most poses from these methods do not precisely align with the target end-effectors. We believe this happens with traditional deep-learning-based approaches due to the use of large networks; since high-frequency details are lost during the forward process.

We observed that latent space optimization methods like HuMoR often encounter difficulties when generating poses for long sequences (e.g., $>60$ poses). This is attributed to the structure and navigation of the latent space, which can lead to the traversal of invalid regions and, subsequently, incorrect results (see our supplemental material for more details). DragPoser mitigates these issues through frame-by-frame optimization, ensuring the continuity of the latent space structure, and utilizing a Temporal Predictor to guide the optimization process.

DragPoser consistently generates high-quality poses across all cases, successfully meeting the end-effector hard constraints. Notably, it excels in reconstructing challenging poses like push-ups \emph{Ours/5} and kicks \emph{Ours/3}, effectively balancing pose accuracy with the fulfillment of hard constraints.

\subsection{Dynamic Constraints Evaluation}
\begin{table}
    \caption{Accuracy of our method for very diverse sparse input scenarios. Each scenario is determined by the number of sensors (3 to 6), their placement, and potential faulty behavior (first column). All sensors have 6 DoFs (position+rotation) unless otherwise noted. We report the joint errors and the end-effector errors, as in Table~\ref{tab:eval:comparison}. 
    \\ *Reported end-effector errors encompass all six sensors; this error is significantly larger compared to more stable setups where only the tracked sensors are evaluated.}
    \scriptsize
    \centering
    \begin{tabular}{lcccc}
        \toprule
        \multirow{3}{*}{Sensors} & \multicolumn{1}{c}{End-Effector} & \multicolumn{3}{c}{Joint Error} \\
         \cmidrule(lr){3-5} \cmidrule(lr){2-2}
        & Pos (\textit{cm}) & Rot (\textit{deg}) & Vel (\textit{cm/s}) & Pos (\textit{cm}) \\
        \midrule
        3 \emph{\tiny{Head+Hands}} & 16.1(11.4) & 22.0(22.3) & 34.0(50.2) & 15.5(15.0) \\
        3 \emph{\tiny{Hip+Hands}} & 1.77(1.31) & 14.1(15.6) & 23.1(36.2) & 9.95(13.7) \\
        4 \emph{\tiny{Hip+Head+Hands}} & 1.12(0.72) & 10.6(12.4) & 20.0(41.5) & 7.86(13.1) \\
        5 \emph{\tiny{Hip+Head+Hands+RFoot}} & 1.17(0.77) & 9.25(10.8) & 15.3(32.8) & 4.95(9.44) \\
        6 \emph{\tiny{Hip+Head+Hands+Feet}} & \multirow{2}{*}{10.2(16.0)*} & \multirow{2}{*}{12.0(14.8)} & \multirow{2}{*}{21.1(59.7)} & \multirow{2}{*}{6.55(11.1)} \\
        \vspace{0.5mm}
        \emph{\tiny{Faulty 1\%}} & & & & \\
        6 \emph{\tiny{Hip+Head+Hands+Feet}} & \multirow{2}{*}{5.85(11.9)*} & \multirow{2}{*}{9.82(12.1)} & \multirow{2}{*}{17.0(50.4)} & \multirow{2}{*}{4.50(8.50)} \\
        \vspace{0.5mm}
        \emph{\tiny{Faulty 0.5\%}} & & & & \\
        6 \emph{\tiny{Hip+Head+Hands+Feet}} & \multirow{2}{*}{1.56(1.51)} & \multirow{2}{*}{25.9(26.2)} & \multirow{2}{*}{13.2(19.6)} & \multirow{2}{*}{5.81(6.71)} \\
        \vspace{0.5mm}
        \emph{\tiny{3 DoF \emph{Position}}} & & & & \\
        6 \emph{\tiny{Hip+Head+Hands+Feet}} & 1.02(0.73) & 7.66(8.71) & 10.7(15.1) & 2.18(2.70) \\
        \bottomrule
    \end{tabular}
    \label{tab:eval:sensors}
\end{table}


One key advantage of DragPoser is its ability to dynamically adapt constraints, accommodating hardware changes or user-specific requirements. Our evaluations demonstrate DragPoser in various sensor configurations, showcasing its flexibility and the potential for customized loss functions to enhance performance in different scenarios as detailed in Table~\ref{tab:eval:sensors}. The baseline configuration employs six 6-DoF, combining positional and rotational data to minimize ambiguity and precisely reconstruct human poses. This setup delivers the highest fidelity in terms of pose quality and end-effector accuracy.

However, in certain applications like motion editing or computer vision, only positional information may be available. When we switch to a configuration with six 3-DoF sensors, DragPoser continues to generate high-quality poses. The \emph{Positional Error} remains similar to the six 6-DoF results of AvatarPoser+, and end-effector accuracy is preserved, although, as expected, there is a notable decrease in rotational precision.

We also explore scenarios with intermittent tracker functionality, simulating faulty sensors by randomly disconnecting one sensor per frame with a probability of 1\% or 0.5\%, reconnecting it after 100 frames. Operating at 60 frames per second, this yields a challenging and unstable input scenario. Despite these conditions, DragPoser robustly reconstructs high-quality poses, achieving metrics on par with those of previous work under stable sensor conditions. It is important to note that the reported end-effector error encompasses all six sensors; thus, due to the frequent disconnections, this error is significantly larger compared to more stable setups where only the tracked sensors are evaluated.

Furthermore, DragPoser's flexibility extends to its ability to work with varying numbers of input sensors. We demonstrate its effective performance with five, four, and three sensors, as previously discussed in comparison with existing work.

Finally, we emphasize that DragPoser's constraints are not limited to the number of sensors. For example, DragPoser could be configured to align a \emph{look-at} vector with the head joint's forward axis, facilitating precise head orientation control. Other examples range from maintaining specified distances between joints to enforcing spatial boundaries. Moreover, hyperparameters like the balance between positional and rotational accuracy or the prioritization of joint accuracy are fully customizable in real time.

\subsection{Performance}

We measured the average execution times of the different components. Note that our code is executed in Python with no special optimizations. On average, the Temporal Predictor forward pass takes around 1.8\,ms per frame. The other components, which run each optimization iteration, have the following running times: the Decoder's forward pass averages 0.3\,ms, the constraints (losses) computation takes around 1.3\,ms (mainly because of the FK), and the backpropagation through the Decoder approximately 2.7\,ms. These three processes are computed on the CPU due to the relatively lightweight Decoder model, while the Temporal Predictor is executed on the GPU.
The maximum execution time per frame can be computed as $1.8 + \text{MAX\_IT}(0.3+1.3+2.7)\,\text{ms}$. By adjusting MAX\_IT, DragPoser is able to adapt to the performance requirements of various application scenarios. Note that this is the maximum execution time per frame; it is usually shorter due to constraints being quickly satisfied.

To assess computational demands, we recorded the average time per frame/pose reconstruction compared to other methods. FinalIK, relying on optimized C\# code, is the fastest at 0.14\,ms. Among data-driven methods, AvatarPoser and SparsePoser lead at 2.15\,ms and 15.6\,ms, respectively, due to their purely forward-pass execution. HuMoR, hampered by sequence-level optimization due to its latent space design, requires 1567\,ms per frame. In contrast, DragPoser, while incorporating optimization, achieves 25.3\,ms per pose, making it suitable for interactive use. Note that we used a maximum iteration count of 40, which is almost never required due to early constraint fulfillment. DragPoser thus bridges the gap between real-time performance and the flexibility of latent space optimization, offering a cost-effective alternative to methods like HuMoR while approaching the speed of forward-pass approaches.

\subsection{Ablation Study}
\label{sec:ablation}
Table~\ref{tab:eval:ablation} presents the results of an ablation study focusing on two critical elements of our proposed method. We first examine the impact of omitting the Continuity Loss $\mathcal{L}_C$ during the training of the Pose Autoencoder. We then evaluate the effectiveness of the Temporal Predictor $\mathcal{T}$ by excluding it from the Pose Optimizer process. All metrics are computed with the six 6-DoF configuration.

\begin{table}
    \scriptsize
    \centering
    \caption{Results of the ablation study. In the first row experiment, the continuity loss $\mathcal{L}_C$ is removed when training the Pose Autoencoder. In the second row experiment the Temporal Predictor $\mathcal{T}$ is not used. We report the joint errors and the end-effector errors.}
    \begin{tabular}{lcccc}
        \toprule
         \multirow{3}{*}{Ablation} & \multicolumn{1}{c}{End-Effector Error} & \multicolumn{3}{c}{Joint Error} \\
         \cmidrule(lr){3-5} \cmidrule(lr){2-2}
         & Pos (\textit{cm}) & Rot (\textit{deg}) & Vel (\textit{cm/s}) & Pos (\textit{cm}) \\
        \midrule
        No $\mathcal{L}_C$ & 1.50(1.13) & 8.39(9.52) & \textbf{9.53(13.3)} & 2.65(2.79) \\
        No $\mathcal{T}$ & 1.36(1.07) & 10.5(12.2) & 13.1(19.0) & 3.05(4.39) \\
        Ours & \textbf{1.02(0.73)} & \textbf{7.66(8.71)} & 10.7(15.1) & \textbf{2.18(2.70)} \\
        \bottomrule
    \end{tabular}
    \label{tab:eval:ablation}
\end{table}



\paragraph*{Continuity Loss}
Our results indicate that including $\mathcal{L}_C$ generally enhances all performance metrics by structuring the latent space more effectively. An exception is observed in the velocity metric; this is attributed to the fact that when the Continuity Loss is used, the optimization process needs fewer iterations to find poses that satisfy the constraints. Consequently, a lower learning rate might be necessary to avoid overshooting the target.

For a practical understanding, we calculated the average number of optimization iterations required to meet the constraint thresholds in two distinct animation scenarios\textemdash a push-up sequence and a dance routine\textemdash setting the maximum number of iterations at 100 per frame. With $\mathcal{L}_C$, the push-up animation required 57 iterations, and the dance 22, on average. Without $\mathcal{L}_C$, these numbers increased significantly to 74 and 73 iterations, respectively.

In summary, the integration of the Continuity Loss $\mathcal{L}_C$ not only improves on finding higher quality poses during the optimization search but also streamlines the pose optimization process, evidenced by a reduced number of iterations required. This leads to a more efficient optimization cycle, reducing the time required to run our method.

\begin{figure*}[ht]
  \includegraphics[width=1\linewidth]{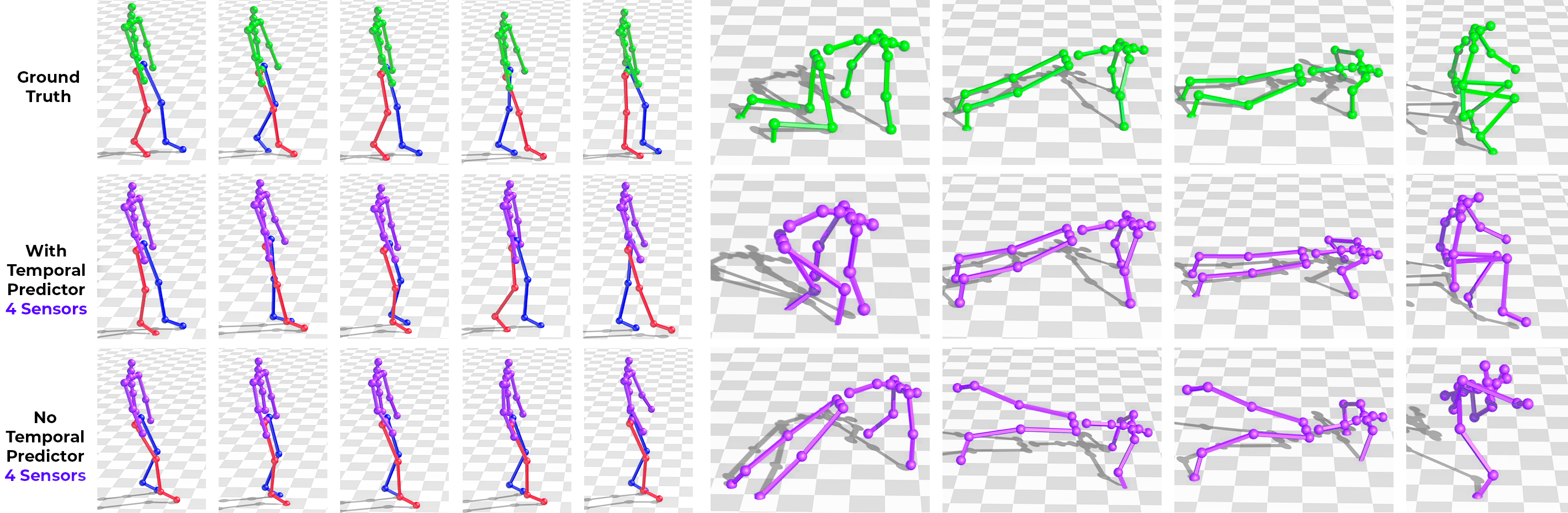}
  \caption{Poses reconstructed using four sparse tracking signals (hip, head and hands) with (middle row) and without (last row) temporal predictor. Ground truth is colored in green. The Temporal Predictor guides the optimization process to produce temporally-coherent motion even in high ambiguity scenarios, thus producing a full walking cycle during motion (left). It also prevents implausible poses by guiding the optimization away from invalid latent space regions (right).}  
  \label{fig:temporal_predictor_comparison}
\end{figure*}

\paragraph*{Temporal Predictor}
The purpose of the Temporal Predictor is twofold. On the one hand, it ensures that predicted latent vectors remain within the manifold of valid poses. On the other hand, it learns the temporality within the latent space to guide the optimization process in high-ambiguity scenarios. In its absence, we observe a significant decline in all metrics, with \emph{Joint Error} notably impacted due to its direct influence on the overall pose quality and not just the optimization of end-effector positioning.

Visual results presented in Figure~\ref{fig:temporal_predictor_comparison} further illustrate the Temporal Predictor's role. Here, the system receives input from only four sensors (hips, head, and hands). With the Temporal Predictor enabled, the system can generate a complete walking cycle; even if it might be out of sync with the ground truth. Without the Temporal Predictor, the character's feet are static, resulting in an unnatural \emph{dragging} motion.

The effectiveness of the Temporal Predictor is even more pronounced in complex motions like push-ups. While the optimization process alone can satisfy end-effector constraints, the Temporal Predictor plays a crucial role in achieving temporally coherent poses. Finally, the last column demonstrates the Temporal Predictor's ability to avoid invalid regions in the latent space: when disabled, the optimization process traverses invalid regions of the pose manifold and ultimately reconstructs an implausible pose.

\section{Conclusions}

In conclusion, our work addresses the critical challenge of achieving high-quality motion reconstruction with a reduced number of input devices, aiming to make motion capture more accessible to a wider audience. 
Our proposed method provides a perfect equilibrium between end-effector accuracy and pose quality.
Through a pose optimization process within a structured latent space, our approach simultaneously achieves superior real-time end-effector position accuracy compared to state-of-the-art IK solutions, and higher pose quality compared to advanced learning approaches like AvatarPoser, SparsePoser and HuMoR.
In addition, our system requires only one-time training on a large human motion dataset, allowing for dynamic constraint definition through losses during runtime. Unlike previous methods limited to specific tracker configurations, our approach can adapt to flexible tracker setups and can handle potential data loss due to occlusions or malfunctions. In such instances, while there may be a slight reduction in pose or end-effector accuracy for certain configurations, the structured latent space and temporal predictor enable our method to generate plausible and smooth animations. Overall, our approach not only expands accessibility to motion capture but also proves to be versatile and effective in handling real-world challenges, showcasing its potential for practical applications.

\paragraph*{Limitations}

Despite the notable achievements of our motion reconstruction system, certain limitations need further consideration. Firstly, in setups with only 3 sensors available, the lack of pelvis/root position tracking has a negative impact on the quality of our motion reconstruction. A promising approach would be to integrate a global position prediction module to overcome this limitation and improve the overall system's performance. Additionally, while our system is trained with various user dimensions and encodes poses using dual quaternions, the lack of comprehensive data specifically focusing on diverse user dimensions means we do not explicitly validate its performance in such scenarios. Another potential limitation arises in instances where the pose optimization process may inadvertently move the latent vector outside the valid pose manifold. Although including the Temporal Predictor effectively encodes valid regions, the possibility of reconstructing invalid poses remains. This is primarily due to the weight used to balance between the Temporal Predictor and the optimization outputs. The lower the weight of the Temporal Predictor, the higher the probability of the optimization process of falling into invalid regions and getting stuck. Furthermore, it is important to note that computing the gradient of the decoder can be time-consuming compared to standard forward prediction. Despite this computational demand, the optimization process typically operates with a few iterations, ensuring relatively fast performance. Finally, DragPoser prioritizes high-frequency detail for accurate end-effector placement, which can sometimes introduce jitter. This can be mitigated by post-processing or by reducing the learning rate of the optimization process. Further research may explore alternative optimization strategies to refine this behavior, such as employing adaptive learning rate methods to control the optimization process better and reduce overshoot.

\paragraph*{Future Work}

Future research directions may involve exploring the integration of various sensor types, such as computer vision, within the dynamic constraints of the pose optimizer. This extension would allow us to have additional sensor data, contributing to a more robust and versatile motion capture system. To further enhance the capabilities of our pose optimizer, we aim to introduce additional constraints, such as look-at functionality or end-effector pose preservation during editing. These improvements could provide more nuanced control over reconstructed motion, meeting diverse user needs and preferences. To refine the latent space structure, we plan to investigate alternative generative approaches. This includes exploring methods that prevent unbounded properties and ensure no empty spaces between valid regions. Such exploration could result in a more compact and well-organized latent space, thus enhancing the system's efficiency and performance. Lastly, despite initially dismissing the idea of directly encoding temporality in the latent vectors due to increased complexity, we recognize it as an intriguing avenue for future research. Currently, we encode this information in an external network. Revisiting this concept and researching efficient strategies to manage the expanded complexity could enhance the latent space representation.

\section{Acknowledgements}
This work has received funding from the European Union’s Horizon 2020 Research and Innovation Programme under HORIZON-CL4-2022-HUMAN-01 grant agreement No 101093159 (XR4ED) and grant agreement No 739578 (RISE), from MCIN/AEI/10.13039/501100011033/FEDER ``A way to make Europe'', UE (PID2021-122136OB-C21), from the Department of Research and Universities of the Government of Catalonia (2021 SGR 01035), and the Government of the Republic of Cyprus through the Deputy Ministry of Research, Innovation and Digital Policy. Jose Luis Ponton was also funded by the Spanish Ministry of Universities (FPU21/01927), and Eduard Pujol by Universitat Politècnica de Catalunya and Banco Santander (FPI-UPC predoctoral grant).


\printbibliography 

@article{Marcard:2017,
author = {von Marcard, Timo and Rosenhahn, Bodo and Black, Michael J. and Pons-Moll, Gerard},
title = {Sparse Inertial Poser: Automatic 3D Human Pose Estimation from Sparse IMUs},
year = {2017},
issue_date = {May 2017},
publisher = {The Eurographics Association and John Wiley & Sons Ltd.},
address = {Chichester, GBR},
volume = {36},
number = {2},
issn = {0167-7055},
doi = {10.1111/cgf.13131},
journal = {Comput. Graph. Forum},
month = may,
pages = {349–360},
numpages = {12}
}

@inproceedings{Agrawal:2023,
 author     = {Agrawal, Dhruv and Guay, Martin and Buhmann, Jakob and Borer, Dominik and W. Sumner, Robert},
 title      = {Pose and Skeleton-aware Neural {IK} for Pose and Motion Editing},
 year       = {2023},
 isbn       = {9798400703157},
 publisher  = {ACM},
 address    = {New York, NY, USA},
 doi        = {10.1145/3610548.3618217},
 booktitle  = {SIGGRAPH Asia 2023 Conference Papers},
 articleno  = {49},
 numpages   = {10},
 location   = {Sydney, Australia},
 series     = {SA'23}
}

@article{Ami-Williams:2023,
title = {Digitizing traditional dances under extreme clothing: The case study of Eyo},
journal = {Journal of Cultural Heritage},
volume = {67},
pages = {145-157},
year = {2024},
issn = {1296-2074},
doi = {https://doi.org/10.1016/j.culher.2024.02.011},
author = {Temi Ami-Williams and Christina-Georgia Serghides and Andreas Aristidou},
}

@article{DanceDB:2019,
	author    	= {Aristidou, Andreas and Shamir, Ariel and Chrysanthou, Yiorgos},
	title 		= {Digital Dance Ethnography: {O}rganizing Large Dance Collections},
	journal 	= {J. Comput. Cult. Herit.},
	issue_date	= {January 2020},
	volume 		= {12},
	number 		= {4},
	month 		= nov,
	year 		= {2019},
	issn 		= {1556-4673},
	articleno 	= {29},
	numpages 	= {27},
	doi 		= {10.1145/3344383},
	acmid 		= {},
	publisher 	= {ACM},
	address 	= {New York, NY, USA},
}

@inproceedings{SOMA:2021,
  title = {{SOMA}: Solving Optical Marker-Based MoCap Automatically},
  author = {Ghorbani, Nima and Black, Michael J.},
  booktitle = {Proc. International Conference on Computer Vision},
  series = {ICCV'21},
  pages = {11117--11126},
  month = oct,
  year = {2021},
  doi = {10.1109/ICCV48922.2021.01093},
  month_numeric = {10}
}

@article{HUMAN4D:2020,
  title={HUMAN4D: A Human-Centric Multimodal Dataset for Motions and Immersive Media},
  author={Chatzitofis, Anargyros and Saroglou, Leonidas and Boutis, Prodromos and Drakoulis, Petros and Zioulis, Nikolaos and Subramanyam, Shishir and Kevelham, Bart and Charbonnier, Caecilia and Cesar, Pablo and Zarpalas, Dimitrios and  Kollias, Stefanos and Daras, Petros},
  journal={IEEE Access},
  volume={8},
  pages={176241--176262},
  year={2020},
  publisher={IEEE},
  doi = {10.21227/xjzb-4y45}
}

@conference{AMASS:2019,
 title      = {{AMASS}: Archive of Motion Capture as Surface Shapes},
 author     = {Mahmood, Naureen and Ghorbani, Nima and Troje, Nikolaus F. and Pons-Moll, Gerard and Black, Michael J.},
 booktitle  = {International Conference on Computer Vision},
 pages      = {5442--5451},
 month      = oct,
 year       = {2019},
 month_numeric = {10},
 doi={10.1109/ICCV.2019.00554}
}

@inproceedings{Loshchilov:2019,
  title         = {Decoupled {{Weight Decay Regularization}}},
  booktitle     = {{{ICLR}} 2019},
  author        = {Loshchilov, Ilya and Hutter, Frank},
  year          = {2019},
  month         = jan,
  eprint        = {1711.05101},
  eprinttype    = {arxiv},
  doi           = {10.48550/arXiv.1711.05101}
}

@article{Huang:2018,
 title      = {Deep Inertial Poser: Learning to Reconstruct Human Pose from Sparse Inertial Measurements in Real Time},
 shorttitle = {Deep Inertial Poser},
 author     = {Huang, Yinghao and Kaufmann, Manuel and Aksan, Emre and Black, Michael J. and Hilliges, Otmar and {Pons-Moll}, Gerard},
 year       = {2018},
 month      = dec,
 journal    = {ACM Transactions on Graphics},
 volume     = {37},
 number     = {6},
 pages      = {185:1--185:15},
 issn       = {0730-0301},
 doi        = {10.1145/3272127.3275108}
}

@inproceedings{Jiang:2022,
 title      = {Transformer {{Inertial Poser}}: {{Real-time Human Motion Reconstruction}} from {{Sparse IMUs}} with {{Simultaneous Terrain Generation}}},
 shorttitle = {Transformer {{Inertial Poser}}},
 booktitle  = {{{SIGGRAPH Asia}} 2022 {{Conference Papers}}},
 author     = {Jiang, Yifeng and Ye, Yuting and Gopinath, Deepak and Won, Jungdam and Winkler, Alexander W. and Liu, C. Karen},
 year       = {2022},
 month      = nov,
 series     = {{{SA}} '22},
 pages      = {1--9},
 publisher  = {{ACM}},
 address    = {{New York, NY, USA}},
 doi        = {10.1145/3550469.3555428},
 isbn       = {978-1-4503-9470-3}
}

@article{Yi:2021,
 title      = {{{TransPose}}: Real-Time {{3D}} Human Translation and Pose Estimation with Six Inertial Sensors},
 shorttitle = {{{TransPose}}},
 author     = {Yi, Xinyu and Zhou, Yuxiao and Xu, Feng},
 year       = {2021},
 month      = jul,
 journal    = {ACM Transactions on Graphics},
 volume     = {40},
 number     = {4},
 pages      = {86:1--86:13},
 issn       = {0730-0301},
 doi        = {10.1145/3450626.3459786}
}

@inproceedings{Yi:2022,
 title      = {Physical {{Inertial Poser}} ({{PIP}}): {{Physics-aware Real-time Human Motion Tracking}} from {{Sparse Inertial Sensors}}},
 shorttitle = {Physical {{Inertial Poser}} ({{PIP}})},
 booktitle  = {2022 {{IEEE}}/{{CVF Conference}} on {{Computer Vision}} and {{Pattern Recognition}} ({{CVPR}})},
 author     = {Yi, Xinyu and Zhou, Yuxiao and Habermann, Marc and Shimada, Soshi and Golyanik, Vladislav and Theobalt, Christian and Xu, Feng},
 year       = {2022},
 month      = jun,
 pages      = {13157--13168},
 publisher  = {{IEEE}},
 address    = {{New Orleans, LA, USA}},
 doi        = {10.1109/CVPR52688.2022.01282},
 isbn       = {978-1-66546-946-3}
}

@article{yi:2023,
  title = {{{EgoLocate}}: {{Real-time Motion Capture}}, {{Localization}}, and {{Mapping}} with {{Sparse Body-mounted Sensors}}},
  shorttitle = {{{EgoLocate}}},
  author = {Yi, Xinyu and Zhou, Yuxiao and Habermann, Marc and Golyanik, Vladislav and Pan, Shaohua and Theobalt, Christian and Xu, Feng},
  year = {2023},
  month = jul,
  journal = {ACM Transactions on Graphics},
  volume = {42},
  number = {4},
  pages = {76:1--76:17},
  issn = {0730-0301},
  doi = {10.1145/3592099}
}

@inproceedings{lee:2024,
  author={Lee, Jiye and Joo, Hanbyul},
  booktitle={2024 IEEE/CVF Conference on Computer Vision and Pattern Recognition (CVPR)}, 
  title={Mocap Everyone Everywhere: Lightweight Motion Capture with Smartwatches and a Head-Mounted Camera}, 
  year={2024},
  volume={},
  number={},
  pages={1091-1100},
  doi={10.1109/CVPR52733.2024.00110}}

@inproceedings{guzov:2021,
  title = {Human {{POSEitioning System}} ({{HPS}}): {{3D Human Pose Estimation}} and {{Self-localization}} in {{Large Scenes}} from {{Body-Mounted Sensors}}},
  shorttitle = {Human {{POSEitioning System}} ({{HPS}})},
  booktitle = {2021 {{IEEE}}/{{CVF Conference}} on {{Computer Vision}} and {{Pattern Recognition}} ({{CVPR}})},
  author = {Guzov, Vladimir and Mir, Aymen and Sattler, Torsten and {Pons-Moll}, Gerard},
  year = {2021},
  month = jun,
  pages = {4316--4327},
  issn = {2575-7075},
  doi = {10.1109/CVPR46437.2021.00430}
}

@article{Ahuja:2021,
author = {Ahuja, Karan and Ofek, Eyal and Gonzalez-Franco, Mar and Holz, Christian and Wilson, Andrew D.},
title = {CoolMoves: User Motion Accentuation in Virtual Reality},
year = {2021},
issue_date = {June 2021},
publisher = {ACM},
address = {New York, NY, USA},
volume = {5},
number = {2},
doi = {10.1145/3463499},
journal = {Proc. ACM Interact. Mob. Wearable Ubiquitous Technol.},
month = jun,
articleno = {52},
numpages = {23}
}

@inproceedings{Dittadi:2021,
  title={Full-Body Motion from a Single Head-Mounted Device: Generating SMPL Poses from Partial Observations},
  author={Dittadi, Andrea and Dziadzio, Sebastian and Cosker, Darren and Lundell, Ben and Cashman, Thomas J and Shotton, Jamie},
  booktitle={Proceedings of the IEEE/CVF International Conference on Computer Vision},
  pages={11687--11697},
  year={2021},
  doi={10.1109/ICCV48922.2021.01148}
}

@inproceedings{Winkler:2022,
 title      = {{{QuestSim}}: {{Human Motion Tracking}} from {{Sparse Sensors}} with {{Simulated Avatars}}},
 shorttitle = {{{QuestSim}}},
 booktitle  = {{{SIGGRAPH Asia}} 2022 {{Conference Papers}}},
 author     = {Winkler, Alexander and Won, Jungdam and Ye, Yuting},
 year       = {2022},
 month      = nov,
 pages      = {1--8},
 publisher  = {{ACM}},
 address    = {{Daegu Republic of Korea}},
 doi        = {10.1145/3550469.3555411},
 isbn       = {978-1-4503-9470-3}
}

@article {Ponton:2022b,
 journal    = {Comp. Graph. Forum},
 title      = {{Combining Motion Matching and Orientation Prediction to Animate Avatars for Consumer-Grade VR Devices}},
 author     = {Ponton, Jose Luis and Yun, Haoran and Andujar, Carlos and Pelechano, Nuria},
 year       = {2022},
 publisher  = {The Eurographics Association and John Wiley & Sons Ltd.},
 ISSN       = {1467-8659},
 DOI        = {10.1111/cgf.14628}
}

@inproceedings{Jiang:2022b,
 title      = {{{AvatarPoser}}: {{Articulated Full-Body Pose Tracking}} from~{{Sparse Motion Sensing}}},
 shorttitle = {{{AvatarPoser}}},
 booktitle  = {Computer {{Vision}} \textendash{} {{ECCV}} 2022},
 author     = {Jiang, Jiaxi and Streli, Paul and Qiu, Huajian and Fender, Andreas and Laich, Larissa and Snape, Patrick and Holz, Christian},
 editor     = {Avidan, Shai and Brostow, Gabriel and Ciss{\'e}, Moustapha and Farinella, Giovanni Maria and Hassner, Tal},
 year       = {2022},
 series     = {Lecture {{Notes}} in {{Computer Science}}},
 pages      = {443--460},
 publisher  = {{Springer Nature Switzerland}},
 address    = {{Cham}},
 doi        = {10.1007/978-3-031-20065-6_26},
 isbn       = {978-3-031-20065-6}
}

@inproceedings{Aliakbarian:2022,
  title = {{{FLAG}}: {{Flow-based 3D Avatar Generation}} from {{Sparse Observations}}},
  shorttitle = {{{FLAG}}},
  booktitle = {2022 {{IEEE}}/{{CVF Conference}} on {{Computer Vision}} and {{Pattern Recognition}} ({{CVPR}})},
  author = {Aliakbarian, Sadegh and Cameron, Pashmina and Bogo, Federica and Fitzgibbon, Andrew and Cashman, Thomas J.},
  year = {2022},
  month = jun,
  pages = {13243--13252},
  publisher = {{IEEE}},
  address = {{New Orleans, LA, USA}},
  doi = {10.1109/CVPR52688.2022.01290},
  isbn = {978-1-66546-946-3}
}

@article{Yang:2021,
  title={Lobstr: Real-time lower-body pose prediction from sparse upper-body tracking signals},
  author={Yang, Dongseok and Kim, Doyeon and Lee, Sung-Hee},
  journal={Comp. Graph. Forum},
  volume={40},
  number={2},
  pages={265--275},
  year={2021},
  publisher={Wiley Online Library},
  doi={10.1111/cgf.142631}
}

@INPROCEEDINGS{Clavet:2016,
  title = {Motion Matching and The Road to Next-Gen Animation},
  author = {Clavet, Simon},
  year = {2016},
  booktitle = {Proc. of the Game Dev. Conference},
  series = {GDC'16}
}

@article{ponton2023sparseposer,
  author = {Ponton, Jose Luis and Yun, Haoran and Aristidou, Andreas and Andujar, Carlos and Pelechano, Nuria},
  title = {SparsePoser: Real-Time Full-Body Motion Reconstruction from Sparse Data},
  year = {2023},
  issue_date = {February 2024},
  publisher = {ACM},
  booktitle = {SIGGRAPH Asia 2023},
  address = {New York, NY, USA},
  volume = {43},
  number = {1},
  issn = {0730-0301},
  doi = {10.1145/3625264},
  journal = {ACM Trans. Graph.},
  month = oct,
  articleno = {5},
  numpages = {14}
}

@article{ye2022,
  title = {{{Neural3Points}}: {L}earning to Generate Physically Realistic Full-body Motion for {V}irtual {R}eality Users},
  shorttitle = {Neural3Points},
  author = {Ye, Yongjing and Liu, Libin and Hu, Lei and Xia, Shihong},
  year = {2022},
  journal = {Comp. Graph. Forum},
  volume = {41},
  number = {8},
  pages = {183-194},
  publisher = {{The Eurographics Association and John Wiley \& Sons Ltd.}},
  issn = {1467-8659},
  doi = {10.1111/cgf.14634}
}

@inproceedings{lee2023,
  title = {{{QuestEnvSim}}: {{Environment-Aware Simulated Motion Tracking}} from {{Sparse Sensors}}},
  shorttitle = {{{QuestEnvSim}}},
  booktitle = {{{ACM SIGGRAPH}} 2023 {{Conference Proceedings}}},
  author = {Lee, Sunmin and Starke, Sebastian and Ye, Yuting and Won, Jungdam and Winkler, Alexander},
  year = {2023},
  month = jul,
  series = {{{SIGGRAPH}} '23},
  pages = {1--9},
  publisher = {{ACM}},
  address = {{New York, NY, USA}},
  doi = {10.1145/3588432.3591504},
  isbn = {9798400701597}
}

@article{milef2023,
  title = {Variational {{Pose Prediction}} with {{Dynamic Sample Selection}} from {{Sparse Tracking Signals}}},
  author = {Milef, Nicholas and Sueda, Shinjiro and Kalantari, Nima Khademi},
  year = {2023},
  journal = {Comp. Graph. Forum},
  volume = {42},
  number = {2},
  pages = {359--369},
  issn = {1467-8659},
  doi = {10.1111/cgf.14767},
  publisher = {Eurographics and John Wiley \& Sons Ltd.}
}

@inproceedings{du2023a,
  title = {Avatars {{Grow Legs}}: {{Generating Smooth Human Motion From Sparse Tracking Inputs With Diffusion Model}}},
  shorttitle = {Avatars {{Grow Legs}}},
  booktitle = {Proceedings of the {{IEEE}}/{{CVF Conference}} on {{Computer Vision}} and {{Pattern Recognition}}},
  author = {Du, Yuming and Kips, Robin and Pumarola, Albert and Starke, Sebastian and Thabet, Ali and Sanakoyeu, Artsiom},
  year = {2023},
  pages = {481--490},
  doi = {10.1109/CVPR52729.2023.00054}
}

@inproceedings{zheng2023,
  title = {Realistic {{Full-Body Tracking}} from {{Sparse Observations}} via {{Joint-Level Modeling}}},
  booktitle = {Proceedings of the {IEEE}/{CVF} International Conference on Computer Vision},
  author = {Zheng, Xiaozheng and Su, Zhuo and Wen, Chao and Xue, Zhou and Jin, Xiaojie},
  year = {2023},
  pages = {14678--14688},
  doi = {10.1109/ICCV51070.2023.01349}
}

@inproceedings{castillo2023,
  title = {{{BoDiffusion}}: {{Diffusing Sparse Observations}} for {{Full-Body Human Motion Synthesis}}},
  shorttitle = {{{BoDiffusion}}},
  booktitle = {2023 {{IEEE}}/{{CVF International Conference}} on {{Computer Vision Workshops}} ({{ICCVW}})},
  author = {Castillo, Angela and Escobar, Maria and Jeanneret, Guillaume and Pumarola, Albert and Arbel{\'a}ez, Pablo and Thabet, Ali and Sanakoyeu, Artsiom},
  year = {2023},
  month = oct,
  pages = {4223--4233},
  issn = {2473-9944},
  doi = {10.1109/ICCVW60793.2023.00456}
}

@article{Aristidou:2018,
author = {Aristidou, Andreas and Lasenby, Joan and Chrysanthou, Yiorgos and Shamir, Ariel},
title = {Inverse Kinematics Techniques in Computer Graphics: A Survey},
journal = {Comp. Graph. Forum},
volume = {37},
number = {6},
pages = {35-58},
doi = {10.1111/cgf.13310},
year = {2018}
}

@article{Caserman:2019,
  title={Real-time body tracking in virtual reality using a Vive tracker},
  author={Caserman, Polona and Garcia-Agundez, Augusto and Konrad, Robert and G{\"o}bel, Stefan and Steinmetz, Ralf},
  journal={Virtual Reality},
  volume={23},
  number={2},
  pages={155--168},
  year={2019},
  publisher={Springer},
  doi={10.1007/s10055-018-0374-z}
}

@article{Grochow:2004,
author = {Grochow, Keith and Martin, Steven L. and Hertzmann, Aaron and Popovi\'{c}, Zoran},
title = {Style-Based Inverse Kinematics},
year = {2004},
issue_date = {August 2004},
publisher = {ACM},
address = {New York, NY, USA},
volume = {23},
number = {3},
issn = {0730-0301},
doi = {10.1145/1015706.1015755},
journal = {ACM Trans. Graph.},
month = aug,
pages = {522–531},
numpages = {10}
}

@ARTICLE{Wu:2011,
  author={Wu, Xiaomao and Tournier, Maxime and Reveret, Lionel},
  journal={IEEE Computer Graphics and Applications}, 
  title={Natural Character Posing from a Large Motion Database}, 
  year={2011},
  volume={31},
  number={3},
  pages={69-77},
  doi={10.1109/MCG.2009.111}
}

@article{Huang:2017,
author = {Huang, Jing and Wang, Qi and Fratarcangeli, Marco and Yan, Ke and Pelachaud, Catherine},
title = {Multi-Variate Gaussian-Based Inverse Kinematics},
journal = {Comp. Graph. Forum},
volume = {36},
number = {8},
pages = {418-428},
doi = {10.1111/cgf.13089},
year = {2017}
}

@article{Victor:2021,
author = {Victor, Léon and Meyer, Alexandre and Bouakaz, Saïda},
title = {Learning-based pose edition for efficient and interactive design},
journal = {Computer Animation and Virtual Worlds},
volume = {32},
number = {3-4},
pages = {e2013},
doi = {10.1002/cav.2013},
year = {2021}
}

@inproceedings{oreshkin2021a,
  title = {{{ProtoRes}}: {{Proto-Residual Network}} for {{Pose Authoring}} via {{Learned Inverse Kinematics}}},
  shorttitle = {{{ProtoRes}}},
  booktitle = {Int. {{Conf.}} on {{Learning Representations}}},
  author = {Oreshkin, Boris N. and Bocquelet, Florent and Harvey, Felix G. and Raitt, Bay and Laflamme, Dominic},
  year = {2021},
  month = oct,
  doi = {10.48550/arXiv.2106.01981}
}

@inproceedings{voleti2022,
  title = {{{SMPL-IK}}: {{Learned Morphology-Aware Inverse Kinematics}} for {{AI Driven Artistic Workflows}}},
  shorttitle = {{{SMPL-IK}}},
  booktitle = {{{SIGGRAPH Asia}} 2022 {{Technical Communications}}},
  author = {Voleti, Vikram and Oreshkin, Boris and Bocquelet, Florent and Harvey, F{\'e}lix and M{\'e}nard, Louis-Simon and Pal, Christopher},
  year = {2022},
  month = nov,
  series = {{{SA}} '22},
  pages = {1--7},
  publisher = {{ACM}},
  address = {{New York, NY, USA}},
  doi = {10.1145/3550340.3564227},
  isbn = {978-1-4503-9465-9}
}

@article{loper2015,
  title = {{{SMPL}}: A Skinned Multi-Person Linear Model},
  shorttitle = {{{SMPL}}},
  author = {Loper, Matthew and Mahmood, Naureen and Romero, Javier and {Pons-Moll}, Gerard and Black, Michael J.},
  year = {2015},
  month = nov,
  journal = {ACM Transactions on Graphics},
  volume = {34},
  number = {6},
  pages = {248:1--248:16},
  issn = {0730-0301},
  doi = {10.1145/2816795.2818013}
}

@article {Andreou:2022,
 author     = {Andreou, Nefeli and Aristidou, Andreas and Chrysanthou, Yiorgos},
 title      = {Pose Representations for Deep Skeletal Animation},
 journal    = {Comp. Graph. Forum},
 volume     = {41},
 number     = {8},
 month      = dec,
 pages      = {},
 year       = {2022},
 publisher  = {The Eurographics Association and John Wiley & Sons Ltd.},
 ISSN       = {1467-8659},
 DOI        = {10.1111/cgf.14632}
}

@article{Aberman:2020,
 title      = {Skeleton-{{Aware Networks}} for {{Deep Motion Retargeting}}},
 author     = {Aberman, Kfir and Li, Peizhuo and Lischinski, Dani and {Sorkine-Hornung}, Olga and {Cohen-Or}, Daniel and Chen, Baoquan},
 year       = {2020},
 month      = aug,
 journal    = {ACM Transactions on Graphics},
 volume     = {39},
 number     = {4},
 issn       = {0730-0301, 1557-7368},
 doi        = {10.1145/3386569.3392462}
}

@inproceedings{Pavllo:2018,
 title      = {{{QuaterNet}}: {{A Quaternion-based Recurrent Model}} for {{Human Motion}}},
 shorttitle = {{{QuaterNet}}},
 author     = {Pavllo, Dario and Grangier, David and Auli, Michael},
 year       = {2018},
 month      = may,
 booktitle  = {British {{Machine Vision Conference}} ({{BMVC}})},
 doi        = {https://arxiv.org/abs/1805.06485}
}

@article{Pavllo:2020,
 title      = {Modeling {{Human Motion}} with {{Quaternion-Based Neural Networks}}},
 author     = {Pavllo, Dario and Feichtenhofer, Christoph and Auli, Michael and Grangier, David},
 year       = {2020},
 month      = apr,
 journal    = {International Journal of Computer Vision},
 volume     = {128},
 number     = {4},
 pages      = {855--872},
 issn       = {1573-1405},
 doi        = {10.1007/s11263-019-01245-6}
}

@article{aristidou:2011,
  title = {{{FABRIK}}: {{A}} Fast, Iterative Solver for the {{Inverse Kinematics}} Problem},
  shorttitle = {{{FABRIK}}},
  author = {Aristidou, Andreas and Lasenby, Joan},
  year = {2011},
  month = sep,
  journal = {Graphical Models},
  volume = {73},
  number = {5},
  pages = {243--260},
  issn = {15240703},
  doi = {10.1016/j.gmod.2011.05.003}
}

@article{kenwright:2012,
  title = {Inverse {{Kinematics}} {\textendash} {{Cyclic Coordinate Descent}} ({{CCD}})},
  author = {Kenwright, Ben},
  year = {2012},
  month = oct,
  journal = {Journal of Graphics Tools},
  volume = {16},
  number = {4},
  pages = {177--217},
  publisher = {{Taylor \& Francis}},
  issn = {2165-347X},
  doi = {10.1080/2165347X.2013.823362}
}

@inproceedings{pan:2023,
  title = {Drag {{Your GAN}}: {{Interactive Point-based Manipulation}} on the {{Generative Image Manifold}}},
  shorttitle = {Drag {{Your GAN}}},
  booktitle = {{{ACM SIGGRAPH}} 2023 {{Conference Proceedings}}},
  author = {Pan, Xingang and Tewari, Ayush and Leimk{\"u}hler, Thomas and Liu, Lingjie and Meka, Abhimitra and Theobalt, Christian},
  year = {2023},
  month = jul,
  pages = {1--11},
  publisher = {ACM},
  address = {{New York, NY, USA}},
  doi = {10.1145/3588432.3591500},
  isbn = {9798400701597}
}

@inproceedings{liu_learning_2022,
	address = {Vancouver BC Canada},
	title = {Learning {Smooth} {Neural} {Functions} via {Lipschitz} {Regularization}},
	isbn = {9781450393379},
	doi = {10.1145/3528233.3530713},
	language = {en},
	booktitle = {Special {Interest} {Group} on CG and {Interactive} {Techniques} {Conference} {Proceedings}},
	publisher = {ACM},
	author = {Liu, Hsueh-Ti Derek and Williams, Francis and Jacobson, Alec and Fidler, Sanja and Litany, Or},
	month = aug,
	year = {2022},
	pages = {1--13},
}

@inproceedings{andrews:2016,
  title = {Real-Time {{Physics-based Motion Capture}} with {{Sparse Sensors}}},
  booktitle = {Proceedings of the 13th {{European Conference}} on {{Visual Media Production}} ({{CVMP}} 2016)},
  author = {Andrews, Sheldon and Huerta, Ivan and Komura, Taku and Sigal, Leonid and Mitchell, Kenny},
  year = {2016},
  month = dec,
  series = {{{CVMP}} '16},
  pages = {1--10},
  publisher = {ACM},
  address = {New York, NY, USA},
  doi = {10.1145/2998559.2998564},
  isbn = {978-1-4503-4744-0}
}

@inproceedings{rempe:2021,
  title = {{{HuMoR}}: {{3D Human Motion Model}} for {{Robust Pose Estimation}}},
  shorttitle = {{{HuMoR}}},
  booktitle = {2021 {{IEEE}}/{{CVF International Conference}} on {{Computer Vision}} ({{ICCV}})},
  author = {Rempe, Davis and Birdal, Tolga and Hertzmann, Aaron and Yang, Jimei and Sridhar, Srinath and Guibas, Leonidas J.},
  year = {2021},
  month = oct,
  pages = {11468--11479},
  issn = {2380-7504},
  doi = {10.1109/ICCV48922.2021.01129},
}

@inproceedings{shi:2023,
  title = {{{PhaseMP}}: {{Robust 3D Pose Estimation}} via {{Phase-conditioned Human Motion Prior}}},
  shorttitle = {{{PhaseMP}}},
  booktitle = {2023 {{IEEE}}/{{CVF International Conference}} on {{Computer Vision}} ({{ICCV}})},
  author = {Shi, Mingyi and Starke, Sebastian and Ye, Yuting and Komura, Taku and Won, Jungdam},
  year = {2023},
  month = oct,
  pages = {14679--14691},
  issn = {2380-7504},
  doi = {10.1109/ICCV51070.2023.01353}
}

@article{starke:2022,
  title = {{{DeepPhase}}: Periodic Autoencoders for Learning Motion Phase Manifolds},
  shorttitle = {{{DeepPhase}}},
  author = {Starke, Sebastian and Mason, Ian and Komura, Taku},
  year = {2022},
  month = jul,
  journal = {ACM Transactions on Graphics},
  volume = {41},
  number = {4},
  pages = {136:1--136:13},
  issn = {0730-0301},
  doi = {10.1145/3528223.3530178}
}

@article{ling:2020,
  title = {Character Controllers Using Motion {{VAEs}}},
  author = {Ling, Hung Yu and Zinno, Fabio and Cheng, George and Van De Panne, Michiel},
  year = {2020},
  month = aug,
  journal = {ACM Transactions on Graphics},
  volume = {39},
  number = {4},
  pages = {40:40:1--40:40:12},
  issn = {0730-0301},
  doi = {10.1145/3386569.3392422}
}

@article{peng:2022,
author = {Peng, Xue Bin and Guo, Yunrong and Halper, Lina and Levine, Sergey and Fidler, Sanja},
title = {ASE: large-scale reusable adversarial skill embeddings for physically simulated characters},
year = {2022},
issue_date = {July 2022},
publisher = {ACM},
address = {New York, NY, USA},
volume = {41},
number = {4},
issn = {0730-0301},
doi = {10.1145/3528223.3530110},
journal = {ACM Trans. Graph.},
month = jul,
articleno = {94},
numpages = {17}
}


\end{document}